\documentclass[aps,prd,notitlepage,showpacs,nofootinbib,superscriptaddress]{revtex4-1}

\usepackage{graphicx}
\usepackage[utf8]{inputenc} 

\usepackage{amsmath}
\usepackage{amssymb}
\usepackage{float}
\usepackage{comment}
\usepackage{slashed}
\usepackage[normalem]{ulem}
\usepackage{booktabs}
\usepackage{array}

\usepackage{caption}
\usepackage{subcaption}
\usepackage{soul}

\usepackage{amsmath}
\usepackage{amssymb}

\usepackage[usenames,dvipsnames]{color}
\usepackage[colorinlistoftodos]{todonotes}
\usepackage[colorlinks=true,citecolor=darkred,urlcolor=darkred, pdfborder={0 0 0}]{hyperref}
\usepackage[normalem]{ulem}

\definecolor{darkred}{rgb}{0.6,0,0}
\usepackage[colorinlistoftodos]{todonotes}

\definecolor{linkcolor}{rgb}{0,0,0.5}

\def\SM{$\mathrm{SU(3)_c \otimes SU(2)_L \otimes U(1)_Y}$ }

\def\TrTrOO{$\mathrm{SU(3)_c \otimes SU(3)_L \otimes U(1)_X \otimes U(1)_N}$ }

 \def\three{\ensuremath{\mathbf{3}}}
 \def\threeS{\ensuremath{\mathbf{3^*}}}

\def\gsim{\raise0.3ex\hbox{$\;>$\kern-0.75em\raise-1.1ex\hbox{$\sim\;$}}}
\def\lsim{\raise0.3ex\hbox{$\;<$\kern-0.75em\raise-1.1ex\hbox{$\sim\;$}}}

\def\SM{$\mathrm{SU(3)_c \otimes SU(2)_L \otimes U(1)_Y}$ }
\newcommand{\sm}{{Standard Model }}

\usepackage{soul}

\definecolor{mightnightblue}{RGB}{25,25,112}

\definecolor{brown}{rgb}{0.59, 0.29, 0.0}

\newcommand {\black} {\color{black}}

\newcommand {\ignore}[1]{}

\def\SM{$\mathrm{SU(3)_c \otimes SU(2)_L \otimes U(1)_Y}$ }
\def\21{$\mathrm{SU(2)_L \otimes U(1)_Y}$}

\def\sm{standard model }

\providecommand{\be}{ \begin{equation} } 
\providecommand{\ee}{ \end{equation} }
\providecommand{\bea}{\begin{eqnarray}}
\providecommand{\eea}{\end{eqnarray}}
\providecommand{\nn}{\nonumber}

\providecommand{\to}{\rightarrow}

\newcommand{\AddrAHEP}{%
  AHEP Group, Institut de F\'{i}sica Corpuscular --
  C.S.I.C./Universitat de Val\`{e}ncia, Parc Cient\'ific de Paterna.\\
 C/ Catedr\'atico Jos\'e Beltr\'an, 2 E-46980 Paterna (Valencia) - SPAIN}

\begin{document}


\title{\boldmath \color{BrickRed} Reloading the Axion in a 3-3-1 setup  }

\author{Alex G. Dias}\email{alex.dias@ufabc.edu.br}
\affiliation{Centro de Ci\^encias Naturais e Humanas, Universidade Federal do ABC,\\ 09210-580, Santo Andr\'e-SP, Brasil}

\author{Julio Leite}\email{julio.leite@ufabc.edu.br}
\affiliation{Centro de Ci\^encias Naturais e Humanas, Universidade Federal do ABC,\\ 09210-580, Santo Andr\'e-SP, Brasil}
\affiliation{\AddrAHEP}

\author{Jos\'{e} W. F. Valle}\email{valle@ific.uv.es}
\affiliation{\AddrAHEP}

\author{Carlos A. Vaquera-Araujo}\email{vaquera@fisica.ugto.mx}
\affiliation{Consejo Nacional de Ciencia y Tecnolog\'ia, Avenida Insurgentes Sur 1582. Colonia Cr\'edito Constructor, Alcald\'ia Benito Ju\'arez, C.P. 03940, Ciudad de M\'exico, M\'exico}
\affiliation{Departamento de F\'isica, DCI, Campus Le\'on, Universidad de
Guanajuato, Loma del Bosque 103, Lomas del Campestre C.P. 37150, Le\'on, Guanajuato, M\'exico}

\begin{abstract}

We generalize the idea of the axion to an extended electroweak gauge symmetry setup.
We propose a minimal axion extension of the Singer-Valle-Schechter (SVS) theory, in which the standard model fits in $\mathrm{SU(3)_L\otimes U(1)_X}$, the number of families results from anomaly cancellation, and the Peccei-Quinn (PQ) solution to the strong-CP problem is implemented.
Neutrino masses arise from a type-I Dirac seesaw mechanism, suppressed by the ratio of SVS and PQ scales, suggesting the existence of new physics at a moderate SVS scale. 
Novel features include an enhanced axion coupling to photons when compared to the DFSZ axion, as well as flavour-changing axion couplings to quarks.

\end{abstract}


\maketitle
\noindent

\section{Introduction}
\label{Sect:intro}

  One of the well-known theoretical loose ends of the standard model consists in understanding the lack of CP violation in the strong interaction.
  A popular way to approach this so-called ``strong CP problem'' is to appeal to the Peccei-Quinn (PQ) mechanism~\cite{Peccei:1977hh}, which leaves a pseudo-Nambu-Goldstone boson, the well-known axion, in the particle spectrum~\cite{Weinberg:1977ma,Wilczek:1977pj}.
  The latter can be realized consistently within the invisible axion approach~\cite{Dine:1981rt,Zhitnitsky:1980tq,Kim:1979if,Shifman:1979if},
  for recent extensive reviews see Refs.~\cite{Sikivie:2020zpn,DiLuzio:2020wdo}.
Another theory challenge is to explain why one has three families of fundamental particles.
A way to approach the latter is to appeal to anomaly cancellation arguments, as in the Singer-Valle-Schechter (SVS) theory~\cite{Singer:1980sw},
or in other subsequent 3-3-1-based proposals~\cite{Frampton:1992wt,Pisano:1991ee}, for a recent short review see~\cite{Byakti:2020ipa}.
Under certain circumstances, this ``anomaly'' mechanism could lead to interesting flavor correlations between rare decays~\cite{Boucenna:2015zwa}.

Last, but not least, there are major physics shortcomings of the standard model, such as the lack of neutrino masses and mixings~\cite{deSalas:2020pgw},
as well as the lack of a viable dark matter candidate~\cite{Bertone:2004pz}. 
While the axion can solve the dark matter problem~\cite{Abbott:1982af,Preskill:1982cy,Dine:1982ah}, it does not come, by itself, accompanied by non-zero neutrino masses and mixings adequate to account for neutrino oscillations~\cite{deSalas:2017kay}.

  There have been recent suggestions on how to relate neutrino mass generation with the strong CP problem~\cite{Carena:2019nnd}.
  This goal can be achieved either assuming Majorana neutrinos~\cite{Dias:2005dn,Dasgupta:2013cwa,Bertolini:2014aia,Ahn:2015pia,Suematsu:2017kcu,Ma:2017zyb,Reig:2018ocz},
  or Dirac-based neutrino mass generation~\cite{Chen:2012baa,Gu:2016hxh,Carvajal:2018ohk,Peinado:2019mrn}.
Indeed, naturally small Dirac neutrino masses may arise effectively, in terms of dimension-five or six operators~\cite{CentellesChulia:2018gwr,CentellesChulia:2018bkz}, as well as from full-fledged, UV-complete, Dirac seesaw theories in which neutrino masses are symmetry-protected. These may be realized either through type-I~\cite{Chulia:2016giq} or type-II seesaw mechanism~\cite{Bonilla:2016zef,Reig:2016ewy}. 

Here we propose a comprehensive SVS-based approach in which all of the above issues appear as closely interconnected.
The field content of the SVS theory~\cite{Singer:1980sw} is extended by the addition of
\begin{itemize}
 \item a single gauge singlet $\sigma$ in the scalar sector, transforming non-trivially under the Peccei-Quinn symmetry,
 \item neutral leptons $S_{aL,R}$, singlets under the 3-3-1 gauge symmetry, but charged under $\mathrm{U(1)_N}$ or $B-L$. 
 \end{itemize}
 
The scalar singlet $\sigma$ field will harbor the axion~\cite{Weinberg:1977ma,Wilczek:1977pj}, as in the invisible DFSZ~\cite{Dine:1981rt,Zhitnitsky:1980tq} or KSVZ~\cite{Kim:1979if} axion models.
On the other hand, the neutral fermions $S_{aL,R}$ will mediate neutrino mass generation through a type-I Dirac seesaw mechanism. 
Our construction differs from all previous implementations of the PQ mechanism within the 3-3-1 framework, for example those suggested in Refs.~\cite{Pal:1994ba,Dias:2003zt,Dias:2003iq,Dias:2018ddy}.
 In particular, our predicted axion couplings to photons and fermions exhibit novel features that we discuss in detail. 

  This paper is organized as follows: in Sec. \ref{sec:model} we sketch the theory setup, field content and quantum numbers under all the symmetries. 
  In Sec.~\ref{sec:scalar} we summarize the scalar sector and symmetry structure,
  while in Secs.~\ref{sec:quark-masses-mixing} and \ref{sec:lept-mass-mixings} we describe the charged fermion Yukawa couplings, and the neutrino mass generation through the Dirac seesaw mechanism, respectively.
  In Section~\ref{sec:axionprop} we summarize the main axion properties.
  Finally, in Sec.~\ref{sec:Conclusions} we present a short discussion and conclude.

\section{Field content and symmetries}
\label{sec:model}
Our model is based on a \TrTrOO extension of the standard model where the symmetry of electromagnetism, $\mathrm{U(1)_Q}$, as well as baryon number minus lepton number symmetry $\mathrm{U(1)_{B-L}}$ remain conserved as residual subgroups after spontaneous breaking takes place. 
Their generators are embedded in the defining symmetry of the model as
\bea\label{QBL}
    Q & = & T_3 - \frac{1}{\sqrt{3}} T_8 + X~,\\
    B-L & = & -\frac{4}{\sqrt{3}} T_8 + N~,
\eea
where $T_3$ and $T_8$ are the diagonal generators of an $\mathrm{SU(3)_L}$ gauge symmetry, while $X$ and $N$ are the generators of the Abelian groups $\mathrm{U(1)_X}$ and $\mathrm{U(1)_N}$, respectively.
There are two possible choices for the Abelian factor $\mathrm{U(1)_N}$. The first one is to keep it global, leading to a conventional 3-3-1 SVS gauge structure~\cite{Singer:1980sw}.
The second one is to promote it to a local symmetry in a fully gauged 3-3-1-1 theory setup~\cite{Dong:2013wca}.
Clearly, a local $\mathrm{U(1)_N}$ symmetry leads to a gauged $B-L$ preserving model, which is viable, provided the associated neutral boson develops an adequate mass.
This can be achieved by the implementation of the Stueckelberg mechanism for $\mathrm{U(1)_N}$ as shown in~\cite{Leite:2020bnb}.
As the important features of our model do not depend on the nature of $B-L$, but only on its preservation, in what follows we adopt an anomaly-free definition of $\mathrm{U(1)_N}$ that can be either gauged or not.

The field content of our model is such that, in the lepton sector, left-handed fields come in the fundamental representation of $\mathrm{SU(3)_L}$, while the right-handed charged leptons appear as $\mathrm{SU(3)_L}$ singlets 
\bea\label{lep}
\psi_{aL}&=&\left( \nu_{aL}, e_{aL},(\nu_{aR})^{c}\right)^{T} \sim \left(\mathbf{1},\three,-\frac{1}{3},-\frac{1}{3}\right),\\
e_{aR} &\sim& \left(\mathbf{1},\mathbf{1},-1,-1\right),\nn 
\eea
where $a=1,2,3$, and the numbers in parentheses represent the field's transformations under the groups $\mathrm{SU(3)_c}$, $\mathrm{SU(3)_L}$, $\mathrm{U(1)_X}$ and $\mathrm{U(1)_N}$, respectively. 
Notice that, in addition to the standard model leptons, the lepton triplets also contain two-component neutral fields, which are identified as (the charge-conjugated) right-handed neutrinos~\cite{Valle:1983dk}. 

For the left-handed quarks, the first two families transform in the anti-fundamental representation of $\mathrm{SU(3)_L}$ and the third in the fundamental representation  
\bea\label{LHqua}
Q_{\alpha L}&=&\left( d_{\alpha L},-u_{\alpha L},D_{\alpha L}\right)^{T}\sim\left(\mathbf{3},\threeS,0,-\frac{1}{3}\right),\\
Q_{3L}&=&\left( u_{3L}, d_{3L},U_{3L}\right)^{T}\sim\left(\mathbf{3},\three,\frac{1}{3},1\right),\nn
\eea
with $\alpha=1,2$.
Such unusual embedding of quark families, in different $\mathrm{SU(3)_L}$ representations, is required to ensure anomaly cancellation~\cite{Singer:1980sw,Frampton:1992wt,Pisano:1991ee}.     
Each triplet component has a right-handed counterpart which is an $\mathrm{SU(3)_L}$ singlet
\bea\label{RHqua}
u_{a R}&\sim&\left(\mathbf{3},\mathbf{1},\frac{2}{3},\frac{1}{3}\right),\quad\quad U_{3R}\sim \left(\mathbf{3},\mathbf{1},\frac{2}{3},\frac{7}{3}\right),\nn\\
d_{a R}&\sim& \left(\mathbf{3},\mathbf{1},-\frac{1}{3},\frac{1}{3}\right), \quad\quad D_{\alpha R}\sim\left(\mathbf{3},\mathbf{1},-\frac{1}{3},-\frac{5}{3}\right).
\eea

In addition to the above fermion fields, already present in the SVS model, we introduce pairs of neutral leptons.
These are vector-like under the gauge symmetries and, therefore, do not contribute to anomaly coefficients
\bea\label{SLR}
S_{aL,R}&\sim& \left(\mathbf{1},\mathbf{1},0,-1\right).
\eea

In the scalar sector, as usual, we consider three fields in the fundamental $\mathrm{SU(3)_L}$ representation 
\bea\label{sca}
\Phi_{1} &=& \left(\phi_{1}^{0},\phi_{1}^{-},\widetilde{\phi}_{1}^{0}\right)^{T}\sim\left(\mathbf{1},\three,-\frac{1}{3},\frac{2}{3}\right),\nn\\
\Phi_2&=&\left(\phi_{2}^{+},\phi_{2}^{0},\widetilde{\phi}_{2}^{+}\right)^{T}\sim\left(\mathbf{1},\three,\frac{2}{3},\frac{2}{3}\right),\\	
\Phi_3&=&\left(\phi_{3}^{0},\phi_{3}^{-},\widetilde{\phi}_{3}^{0}\right)^{T}\sim \left(\mathbf{1},\three,-\frac{1}{3},-\frac{4}{3}\right).\nn
\eea
Finally, we introduce a scalar gauge singlet
\bea\label{sig}
\sigma &\sim& \left(\mathbf{1},\mathbf{1},0,0\right).
\eea

Besides the defining symmetries of the model, we assume that the classical Lagrangian displays a global Peccei-Quinn symmetry.
In Table \ref{tab1}, we present a summary of the transformation properties of fermions and scalars.
We have parameterized the most general $\mathrm{U(1)_{PQ}}$ charge assignment in terms of the charges of $Q_{\alpha L}, \Phi_1, \Phi_3$ and $\sigma$.

The anomaly relevant for the Peccei-Quinn symmetry is the QCD anomaly, $[\mathrm{SU(3)_c}]^2\times \mathrm{U(1)_{PQ}}$, with coefficient 
\begin{eqnarray}\label{Cag}
 C_{ag}&=& \sum_{quarks} \left(PQ_{q_L}-PQ_{q_R}\right) \\
 &=& 6\, PQ_{Q_{\alpha L}}+3\, PQ_{Q_{3 L}} - \left(3\, PQ_{u_{aR}}+3\, PQ_{d_{aR}}+ PQ_{U_{3R}}+2\, PQ_{D_{\alpha R}} \right)= PQ_{\sigma}\,.\nonumber
\end{eqnarray}
Thus in order to solve the strong-CP problem through the Peccei-Quinn mechanism, we require $PQ_{\sigma}\neq 0$.
In Section~\ref{sec:axionprop} we discuss in detail the properties of the axion in our model.

\begin{table}[h!]
	\begin{centering}
     {\renewcommand{\arraystretch}{1.5}
		\begin{tabular}{ccccc}
            \hline
			\toprule 
			\,\,Field\,\, &\,\,\,\,\,\, 3-3-1-1 rep\,\,\,\,\,\, &\,\, \,\,\,\,\,\,$B-L$\,\,\,\,\,\,\,\, &\,\,$\mathrm{U(1)_{PQ}}$\,\,\,
			\tabularnewline
			\hline\hline
			\midrule
			$\psi_{aL}$ & $\left(\mathbf{1},\three,-\frac{1}{3},-\frac{1}{3}\right)$  & $\left(-1,-1,+1\right)^{T}$&\,\,\,\,$\frac{1}{2}(-PQ_{\sigma}+PQ_{\Phi_1}+PQ_{\Phi_3})$\,\, \tabularnewline
			$e_{aR}$ & $\left(\mathbf{1},\mathbf{1},-1,-1\right)$ & $-1$ &$\frac{1}{2}(PQ_{\sigma}+3PQ_{\Phi_1}+3PQ_{\Phi_3})$ \tabularnewline
			$Q_{\alpha L}$ & $\left(\mathbf{3},\threeS,0,-\frac{1}{3}\right)$ & $\left(\frac{1}{3},\frac{1}{3},-\frac{5}{3}\right)^{T}$ & $PQ_{Q_{\alpha L}}$\tabularnewline
			$Q_{3L}$ & $\left(\mathbf{3},\three,\frac{1}{3},1\right)$  & $\left(\frac{1}{3},\frac{1}{3},\frac{7}{3}\right)^{T}$& $PQ_{Q_{\alpha L}}-PQ_{\sigma}-PQ_{\Phi_3}$ \tabularnewline
			$u_{a R}$ & $\left(\mathbf{3},\mathbf{1},\frac{2}{3},\frac{1}{3}\right)$ & $\frac{1}{3}$ &\,\,\,$PQ_{Q_{\alpha L}}-(PQ_{\sigma}+PQ_{\Phi_1}+PQ_{\Phi_3})$\tabularnewline
			$ U_{3R}$ & $\left(\mathbf{3},\mathbf{1},\frac{2}{3},\frac{7}{3}\right)$ & $\frac{7}{3}$ &$PQ_{Q_{\alpha L}}-PQ_{\sigma}-2PQ_{\Phi_3}$\tabularnewline
			$d_{a R}$ & $\left(\mathbf{3},\mathbf{1},-\frac{1}{3},\frac{1}{3}\right)$ & $\frac{1}{3}$ &$PQ_{Q_{\alpha L}}+PQ_{\Phi_1}$\tabularnewline
            $D_{\alpha R}$ & $\left(\mathbf{3},\mathbf{1},-\frac{1}{3},-\frac{5}{3}\right)$  & $-\frac{5}{3}$ & $PQ_{Q_{\alpha L}}+PQ_{\Phi_3}$\tabularnewline
            \hline
            $S_{aL}$ & $\left(\mathbf{1},\mathbf{1},0,-1\right)$ & $-1$ &$\frac{1}{2}(PQ_{\sigma}-PQ_{\Phi_1}+PQ_{\Phi_3})$
           \tabularnewline
            $S_{aR}$ & $\left(\mathbf{1},\mathbf{1},0,-1\right)$ & $-1$ & $\frac{1}{2}(-PQ_{\sigma}-PQ_{\Phi_1}+PQ_{\Phi_3})$
            \tabularnewline
            \hline \hline
			$\Phi_{1}$ & $\left(\mathbf{1},\three,-\frac{1}{3},\frac{2}{3}\right)$  & $\left(0,0,2\right)^{T}$ &$PQ_{\Phi_1}$\tabularnewline
			$\Phi_2$ & $\left(\mathbf{1},\three,\frac{2}{3},\frac{2}{3}\right)$ & $\left(0,0,2\right)^{T}$  & $-(PQ_{\sigma}+PQ_{\Phi_1}+PQ_{\Phi_3})$\tabularnewline
			$\Phi_3$ & $\left(\mathbf{1},\three,-\frac{1}{3},-\frac{4}{3}\right)$ & $\left(-2,-2,0\right)^{T}$ &$PQ_{\Phi_3}$\tabularnewline
			\hline
			\,\,\,$\sigma$\,\,\, & \,\,\,$\left(\mathbf{1},\mathbf{1},0,0\right)$\,\,\,& \,\,\,$0$\,\,\, & $PQ_{\sigma}$
\tabularnewline
			\bottomrule
			\hline
		\end{tabular}}
		\par\end{centering}
		\caption{Field content and symmetry transformations}\label{tab1}
\end{table}

\section{Scalar Sector and symmetry structure}
\label{sec:scalar}

The scalar potential associated to the field content and symmetry properties shown in Table \ref{tab1}, can be divided in two parts $V=V_{1}+V_{2}$.
The first contribution only contains the usual three triplets of the SVS theory,
\begin{eqnarray}\label{V1}
V_1 = \,
 \sum_{i=1}^{3}\left[\mu_i^2 \Phi^{\dagger}_i\Phi_i+\lambda_{i}(\Phi^{\dagger}_i\Phi_i)^2\right]+\sum_{i<j}^3\left[\lambda_{ij}(\Phi^{\dagger}_i\Phi_i)(\Phi^{\dagger}_j\Phi_j)+\tilde{\lambda}_{ij}(\Phi^{\dagger}_i\Phi_j)(\Phi^{\dagger}_j\Phi_i)
\right]\,,
\end{eqnarray}
which are decomposed as
\begin{equation}
\label{scalar3plets3}
\Phi_1=\left(
\begin{array}{c}
\frac{v_1+s_1+i a_1}{\sqrt{2}}\\
\phi^{-}_{1}\\
\widetilde{\phi}^{0}_{1}
\end{array}
\right),\qquad
\Phi_2=\left(
\begin{array}{c}
\phi_2^{+}\\
\frac{v_2+s_2+i a_2}{\sqrt{2}}\\
\widetilde{\phi}_2^+
\end{array}
\right),\qquad
\Phi_3=\left(
\begin{array}{c}
 \phi_3^{0} \\
 \phi_3^- \\
 \frac{w+s_3+i a_3}{\sqrt{2}} \\
\end{array}
\right).
\end{equation}

On the other hand, $V_2$ includes all possible terms involving the scalar singlet $\sigma$ 
\begin{eqnarray}\label{V2}
V_2 &=& \,
 \mu_\sigma^2 \sigma^*\sigma+\lambda_\sigma(\sigma^*\sigma)^2+ \sum_{i}^3 \lambda_{i\sigma}(\Phi^{\dagger}_i\Phi_i)(\sigma^*\sigma) -\left(\lambda_A\, \sigma\,\Phi_1\Phi_2\Phi_3+\mathrm{h.c.}\right).
\end{eqnarray}
where the scalar singlet is written as
\begin{equation}
\sigma= \frac{v_\sigma+s_\sigma+i a_\sigma}{\sqrt{2}}\,.
\end{equation}

The above vacuum alignment, satisfying the hierarchies $v_\sigma^2\gg w^2\gg v_1^2+v_2^2\equiv v^2_{EW}$, ensures that the $B-L$ symmetry is conserved.
From the scalar potential and field decomposition, we extract the following extremum conditions
\begin{eqnarray}
 v_1 \left( 2 \mu_1^2 + 2 \lambda_{1} v_1^2 + \lambda_{12} v_2^2  + \lambda_{13} w^2   + \lambda_{1\sigma} v_\sigma^2 \right) &=& \lambda_A v_2 w  v_\sigma\,,\\
 v_2 \left( 2 \mu_2^2 +  2 \lambda_2 v_2^2  + \lambda_{12} v_1^2  + \lambda_{23} w^2   + \lambda_{2\sigma} v_\sigma^2   \right) &=& \lambda_A v_1 w v_\sigma \,,\nonumber\\
 w \left( 2 \mu_3^2 + \lambda_{13} v_1^2  + \lambda_{23} v_2^2 + 2 \lambda_{3} w^2 + \lambda_{3\sigma} v_\sigma^2  \right) &=& \lambda_A v_1 v_2 v_\sigma\,,\nonumber\\
 v_\sigma \left( 2 \mu_\sigma^2 + \lambda_{1\sigma} v_1^2 + \lambda_{2\sigma} v_2^2 + \lambda_{3\sigma} w^2 + 2 \lambda_{\sigma}  v_\sigma^2 \right) &=& \lambda_A v_1 v_2 w\,,\nonumber
\end{eqnarray}
which we solve simultaneously for the dimensionful constants $\mu_1,\mu_2, \mu_3$ and $\mu_\sigma$.

We now calculate the tree-level scalar spectrum. First we consider the CP-odd scalars. When grouped together in the basis $(a_1, a_2, a_3, a_\sigma )$,
these states share the squared mass matrix
\begin{eqnarray}
M_a^2 = \frac{\lambda_A}{2}\left(
\begin{array}{cccc}
 \frac{v_2 w v_\sigma}{v_1} &  w v_\sigma & v_2 v_\sigma & v_2 w \\
  w v_\sigma & \frac{v_1 w v_\sigma}{v_2} & v_1 v_\sigma & v_1 w \\
 v_2 v_\sigma & v_1 v_\sigma & \frac{v_1 v_2 v_\sigma}{w} & v_1 v_2 \\
 v_2 w & v_1 w & v_1 v_2 & \frac{v_1 v_2 w}{v_\sigma} \\
\end{array}
\right)\,.
\end{eqnarray}
By diagonalizing $M_a^2$, we find that only one state 
\begin{equation}\label{A}
 A = \frac{1}{\sqrt{N_A}}\left[  v_2 w v_\sigma\, a_1 + v_1 w v_\sigma\, a_2 + v_1 v_2 v_\sigma \,a_3 +  v_1v_2w\,a_\sigma \right]\,,
\end{equation}
where 
\begin{equation}
 N_A=v_1^2 v_2^2 w^2+ v_\sigma^2( v_1^2 v_2^2  + v^2_{EW} w^2)\,,
\end{equation}
gets a large mass after spontaneous symmetry breaking,
\begin{equation}
 m_A^2 = \lambda_A \frac{ v_1^2 v_2^2 w^2+ v_\sigma^2( v_1^2 v_2^2  + v^2_{EW} w^2)}{2 v_1 v_2 w v_\sigma}  \,.
\end{equation}  

Two other mass eigenstates, $G_1$ and $G_2$, are would-be Goldstone bosons. They are absorbed by the neutral vector bosons of $\mathrm{SU(3)_L}$ through the Higgs mechanism. They do not have a component along $\sigma$.

Finally, the last (apparently) massless field is actually the axion associated with the spontaneous breaking of the anomalous $\mathrm{U(1)_{PQ}}$ symmetry and is given by
\begin{equation}\label{axion}
 a = \frac{1}{\sqrt{N_a}}\left[ -v_1 v_2^2 w^2\, a_1 - v_2 v_1^2 w^2\, a_2 - w v_1^2 v_2^2\, a_3 + v_\sigma(v_1^2 v_2^2+v_{EW}^2 w^2 )\,a_\sigma \right]\,,
\end{equation}
where the normalization constant $N_a$ is given by
\begin{equation}\label{Na}
 N_a=\left(v_1^2 v_2^2 + v_{EW}^2 w^2\right)N_A\,.
\end{equation}
One sees that, in the limit of interest, $v_\sigma\gg w \gg v_1,v_2$, the axion is mainly the imaginary part of $\sigma$.\\[-.2cm]

Turning now to the CP-even scalars, in the basis $(s_1, s_2, s_3, s_\sigma )$, the relevant squared mass matrix is given by
\begin{eqnarray}
M_s^2 = \frac{1}{2}\left(
\begin{array}{cccc}
 2 \lambda_{1} v_1^2+\frac{\lambda_A v_2 w v_\sigma }{ v_1} & 2\lambda_{12} v_1 v_2 -\lambda_A w v_\sigma & 2\lambda_{13} v_1 w  - \lambda_A v_2 v_\sigma & 2\lambda_{1 \sigma} v_1 v_\sigma -\lambda_A v_2 w \\
 2\lambda_{12} v_1 v_2-\lambda_A w v_\sigma & 2 \lambda_{2} v_2^2 +\frac{\lambda_A v_1 v_\sigma w}{ v_2} &  2v_2 w \lambda_{23}-\lambda_A v_1 v_\sigma  &  2\lambda_{2 \sigma} v_2 v_\sigma -\lambda_A v_1 w\\
 2 \lambda_{13} v_1 w -\lambda_A v_2 v_\sigma & 2\lambda_{23}v_2 w -\lambda_A v_1 v_\sigma &2 \lambda_{3} w^2 + \frac{\lambda_A v_1 v_2 v_\sigma}{ w} & 2\lambda_{3 \sigma}w v_\sigma -\lambda_A v_1 v_2 \\
 2\lambda_{1 \sigma} v_1 v_\sigma-\lambda_A v_2 w & 2\lambda_{2 \sigma} v_2 v_\sigma -\lambda_A v_1 w & 2\lambda_{3 \sigma} w v_\sigma  - \lambda_A v_1 v_2 &4 \lambda_\sigma  v_\sigma^2 + \frac{\lambda_A v_1 v_2 w}{ v_\sigma} \\
\end{array}
\right)\,.
\end{eqnarray}
In general, the matrix above leads to four non-vanishing eigenvalues, associated to four massive scalar bosons, $H_1$, $H_2$, $H_3$ and $H_4$.
For $v_\sigma= 10^{12}$ GeV, $w= 10^{4}$ GeV, and $\sqrt{v_1^2+v_2^2}= 246$ GeV, the heavier state is $H_4 \simeq s_\sigma$, which becomes much heavier than the others,
$m_{H_4}^2 \simeq 2 \lambda_\sigma v_\sigma^2 $, and hence decouples from the rest. 
The lighter state is identified with the $125$ GeV Higgs boson, $H_1 \equiv h$. The remaining states, $H_2$ and $H_3$, get masses around the SVS scale $w$.\\[-.2cm]

In addition to the neutral scalars presented above, the model counts with the complex neutral fields $\widetilde{\phi}_1^0$ and $\phi_3^{0}$ which have opposite $B-L$ charge, and when grouped in the basis $(\widetilde{\phi}_1^0,\phi_3^{0*} )$, share the following squared mass matrix
\begin{eqnarray}
M_{\phi^0}^2 = \frac{1}{2}\left(
\begin{array}{cc}
 w \left(\tilde{\lambda}_{13} w +\frac{\lambda_A v_2 v_\sigma}{v_1}\right) & \lambda_Av_2 v_\sigma + \tilde{\lambda}_{13} v_1 w  \\ \lambda_A v_2 v_\sigma + \tilde{\lambda}_{13} v_1 w  & v_1 \left(\tilde{\lambda}_{13} v_1 +\frac{\lambda_A v_2 v_\sigma}{w}\right) \\
\end{array}
\right)\,.
\end{eqnarray}
In the mass basis, only one of the states appears in the physical spectrum
\begin{eqnarray}
\varphi^0=\frac{w \widetilde{\phi}_1^0+v_1 \phi_3^{0*}}{\sqrt{v_1^2+w^2}},
\end{eqnarray}
and has a heavy squared mass
\begin{equation}
m_{\varphi^0}^2=\frac{(v_1^2+w^2) (\tilde{\lambda}_{13} v_1 w+\lambda_A v_2 v_\sigma)}{2 v_1 w}\,.
\end{equation}
The other state $G_3$, orthogonal to $\varphi^0$, is massless and absorbed by the gauge sector.\\[-.2cm]

Finally, writing the charged scalars in the basis $(\phi_2^\pm,\phi_1^\pm,\widetilde{\phi}_2^\pm,\phi_3^\pm )$, we find the squared mass matrix
\begin{eqnarray}
M_\pm^2 = \frac{1}{2}\left(
\begin{array}{cccc}
 v_1 \left(\tilde{\lambda}_{12}v_1 +\frac{\lambda_A w v_\sigma }{v_2}\right) & \lambda_A w v_\sigma + \tilde{\lambda}_{12} v_1 v_2  & 0 & 0 \\
 \lambda_A w v_\sigma + \tilde{\lambda}_{12} v_1 v_2 & v_2 \left(\tilde{\lambda}_{12} v_2 +\frac{\lambda_A w v_\sigma }{v_1v_2w}\right) & 0 & 0 \\
 0 & 0 & w \left( \tilde{\lambda}_{23} w +\frac{\lambda_A v_1 v_\sigma}{v_2}\right) & \lambda_A v_1 v_\sigma + \tilde{\lambda}_{23} v_2 w  \\
 0 & 0 & \lambda_A v_1 v_\sigma + \tilde{\lambda}_{23} v_2 w & v_2 \left( \tilde{\lambda}_{23}v_2 +\frac{\lambda_A v_1 v_\sigma}{w}\right) \\
\end{array}
\right)\,.
\end{eqnarray}
As expected, charged fields with different $B-L$ charges do not mix.
Diagonalizing the matrix above, we find two heavy charged scalar fields
\begin{equation}
H^{\pm}_1= \frac{v_1 \phi^{\pm}_2+v_2 \phi^{\pm}_1}{\sqrt{v_1^2+v_2^2}}\,,\quad\quad H^{\pm}_2=\frac{w\widetilde{\phi}_2^{\pm}+v_2 \phi^{\pm}_3  }{\sqrt{v_2^2 + w^2}}\,,
\end{equation}
whose masses are
\begin{eqnarray}
m_{H_1^\pm}^2&=&\frac{\left(v_1^2+v_2^2\right) (\tilde{\lambda}_{12}v_1 v_2 +\lambda_A w v_\sigma)}{2 v_1 v_2}\,,\\
m_{H_2^\pm}^2&=&\frac{\left(v_2^2+w^2\right) (\tilde{\lambda}_{23}v_2 w +\lambda_A v_1 v_\sigma)}{2 v_2 w}\, ,\nonumber
\end{eqnarray}
while the other two  massless states, $G_4^\pm$ and $G_5^\pm$, are absorbed by the charged gauge boson sector.

\section{Quark masses and mixing}
\label{sec:quark-masses-mixing}

The allowed Yukawa interactions for the quarks are given as
\begin{eqnarray}\label{lagYq}
 -\mathcal{L}_{\rm Yq} &=& \,
y^{u}_{\alpha a}\, \overline{Q_L^{\alpha}}\,\Phi_2^{*} \, u^{a}_R   +  
y^{u}_{3a}\, \overline{Q_L^{3}}\, \Phi_1 \,u^{a}_R  
+ y^{d}_{3a} \, \overline{Q_L^{3}}\, \Phi_2 \, d^{a}_R  + 
y^{d}_{\alpha a} \, \overline{Q_L^{\alpha}}\,\Phi_1^{*} \,d^{a}_R  \nonumber \\
&+& 
 y^{U}_{33}\, \overline{Q_L^{3}} \,\Phi_3  \,U^3_R
  +  y^{D}_{\alpha\beta} \, \overline{Q_L^{\alpha}}\,\Phi_3^{*}\,D_R^{\beta}  
 + \mathrm{h.c.}
\end{eqnarray}
When the scalar fields acquire vacuum expectation values (vevs), the up-type quarks get the following mass matrix: 
\begin{equation}\label{uqmass}
M_{u}=\frac{1}{\sqrt{2}}\left(
\begin{array}{cccc}
 -v_2 y^u_{11} & -v_2  y^u_{12} &  -v_2  y^u_{13} & 0 \\
 -v_2 y^u_{21} &  -v_2 y^u_{22} &  -v_2  y^u_{23} & 0 \\
 v_1 y^u_{31} &  v_1  y^u_{32} &  v_1  y^u_{33} & 0 \\
0 & 0 & 0 & w y^{U}_{33} \\
\end{array}
\right)=\left(
\begin{array}{cccc}
 m^u_{3\times 3} &  0_{3\times 1} \\
0_{1\times 3} & \frac{w y^{U}_{33}}{\sqrt{2}} \\
\end{array}
\right),
\end{equation}
in the basis $(u_{a},U_3)$.
The $3\times 3$ mass matrix associated with the standard up-type quarks is diagonalized by rotating the left and right fields to the mass basis according to
  $u_{L,R}\to U^u_{L,R}\, u^\prime_{L,R}$, leading to $ \mbox{diag}(m_u, m_c, m_t)=(U_{L}^{u})^\dagger m_{3\times 3}^u U_{R}^{u}$.

On the other hand, the down-type quarks, in the basis $(d_{a},D_{\alpha})$, acquire the mass matrix
\begin{equation}\label{dqmass}
\begin{split}
M_{d}&=\frac{1}{\sqrt{2}}\left(
\begin{array}{ccccc}
 v_1 y^d_{11} & v_1  y^d_{12} &  v_1  y^d_{13} & 0 &0\\
  v_1 y^d_{21} &   v_1 y^d_{22} &  v_1  y^d_{23} & 0 &0\\
  v_2  y^d_{31} &   v_2  y^d_{32} &  v_2  y^d_{33} & 0 &0\\
0 & 0 & 0 &  w y_{11}^{D} & w y_{12}^{D} \\
0 & 0 & 0 &  w y_{12}^{D} &  w y_{22}^{D} \\
\end{array}
\right)=\left(
\begin{array}{cccc}
 m^d_{3\times 3} &  0_{3\times 2} \\
0_{2\times 3} & m^D_{2\times 2} \\
\end{array}
\right).
\end{split}
\end{equation}
As in the previous case, the mass matrix of the standard down-type quarks, $m^d_{3\times 3}$, is diagonalized by rotating the flavor states to the mass basis:
  $d_{L,R}\to U^d_{L,R}\, d^\prime_{L,R}$, so as to obtain $ \mbox{diag}(m_d, m_s, m_b)=(U_{L}^{d})^\dagger m_{3\times 3}^d U_{R}^{d}$.

Notice that the conservation of the $B-L$ symmetry ensures that the exotic quarks do not mix with the standard ones, making sure that the Cabibbo-Kobayashi-Maskawa matrix describing light quark mixing,
and defined as
\begin{equation}
  \label{eq:CKM}
  V_{CKM}=(U_{L}^u)^\dagger U_{L}^d,
\end{equation}
is strictly unitary, as in the standard model. \\[-.2cm]

\section{Lepton masses and mixing}
\label{sec:lept-mass-mixings}

On the other hand, turning to the lepton sector, we have the Yukawa Lagrangian
\begin{eqnarray}\label{lagYl}
 -\mathcal{L}_{\rm Yl} &=& \,
 y^{e}_{ab}\,\overline{\psi_{aL} }  \,\Phi_2 e_{bR}  + y^{\nu_1}_{ab} \,\overline{ \psi_{aL}}\,\Phi_1 \,S_{bR} + y^{\nu_2}_{ab} \,\overline{ \psi_{aL}}\,\Phi_3 \, (S_{bL})^c +
 y^{S}_{ab}\, \overline{S_{aL}} \,\sigma\, S_{bR}  + \mathrm{h.c.},
\end{eqnarray}
so that the charged lepton masses can be obtained simply as
\begin{equation}
 M_e = \frac{y^e v_2}{\sqrt{2}}\,,
\end{equation}
where the family indices have been omitted. Again here the mass matrix is diagonalized as $\mbox{diag}(m_e, m_\mu, m_\tau)=(U_{L}^{e})^\dagger M_e U_{R}^{e}$,
where $U_{L,R}^{e}$ are the unitary matrices connecting the left/right flavor, $e_{L,R}$, and mass eigenstates, $e^\prime_{L,R}$.\\[-.2cm]

\begin{figure}[htbp]
\begin{center}
\includegraphics[scale=1.2]{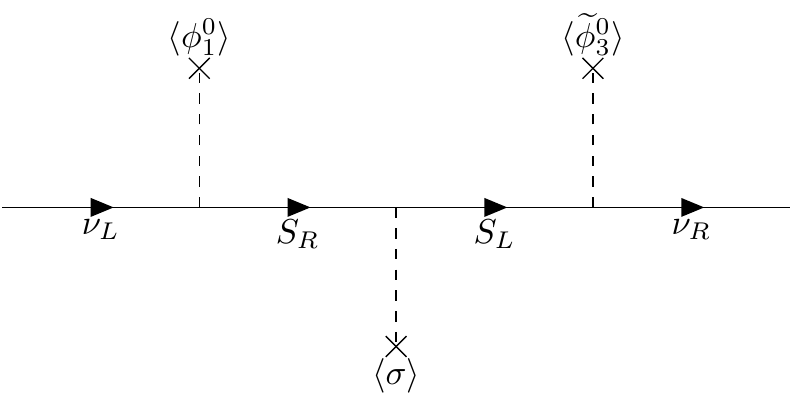}
\caption{Type-I Dirac seesaw mechanism for neutrino masses}
\label{DiracSeesawDia}\label{fig}
\end{center}
\end{figure}

We now turn to the structure of neutrino masses and mixing. Here we first note that the PQ symmetry forbids the term $\overline{\psi_L} \Phi_2^* (\psi_L)^c$ which would generate an unsuppressed
Dirac neutrino mass.
As a result, neutrino masses are generated via the type-I Dirac seesaw mechanism, illustrated in Fig. \ref{DiracSeesawDia}.
In the basis $N=(\nu, S)$, we can write neutral mass term $\overline{N_L} M_{Dirac} N_R$ in terms of the seesaw-type-I matrix, 
\begin{eqnarray}\label{ssm}
 M_{Dirac} =\frac{1}{\sqrt{2}}\begin{pmatrix} 
       0 & y^{\nu_1} v_1 \\ (y^{\nu_2})^T w & y^{S} v_\sigma
      \end{pmatrix},
\end{eqnarray}
where ``Dirac'' indicates that all terms are Dirac-type. This matrix can be written in a diagonal form as $\mbox{diag}(m^N_n)=(U_L^N)^\dagger M_D(U_R^N)$, with $n=1,...,6$, once the
chiral flavor fields are rotated to the mass basis through the unitary transformations $N_{L,R} \to U_{L,R}^N N_{L,R}^\prime$.
The full Dirac seesaw expansion formula is readily obtained from the method in Ref.~\cite{Schechter:1981cv}, though here it suffices for us to keep just the first order term,
\begin{equation} \label{ssm2}
m_\nu^D \simeq \frac{ y^{\nu_1} (y^S)^{-1} (y^{\nu_2})^T }{\sqrt{2}}   \frac{v_1 w}{v_\sigma}\,.
\end{equation}
One sees how the small active neutrino masses result from the suppression by the large seesaw mediator mass, which is identified to lie at the Peccei-Quinn scale.
Choosing $v_\sigma \simeq v_{PQ}$ suggests the existence of new physics at a \textit{lower} scale $w$, characterizing the extended electroweak gauge sector of the SVS theory.
For example with $v_1 = 10^2$ GeV, $w= 10^{4}$ GeV and $v_\sigma = 10^{12}$ GeV, sub-eV neutrino masses ($0.1$ eV) are obtained for reasonable Yukawa couplings $y^{\nu_{1,2}}\sim 10^{-2}$ and $y^{S}\sim 1$.

Likewise, the lepton mixing matrix describing neutrino oscillations arises as
\begin{equation}
  \label{eq:lepton-mixing}
   V_{LEP}=(U_{L}^e)^\dagger U_{L}^\nu,
\end{equation}
where the charged lepton piece is completely standard, while the neutral piece involves also the mediator fermions.
Since these lie at the Peccei-Quinn scale, the mixing matrix of the light neutrinos is nearly unitary, and conveniently described as in the case of quarks.
Indeed, since neutrinos are Dirac-type, the would-be Majorana phases are not physical and can be removed by field redefinition~\cite{Schechter:1980gr,Rodejohann:2011vc}.
No family symmetry is assumed, hence the lepton mixing matrix is totally arbitrary and chosen to fit the observed pattern of neutrino oscillations~\cite{deSalas:2017kay}.\\[-.2cm]

\section{Basic axion properties}
\label{sec:axionprop}

Having presented the scalar and fermion spectra, we now turn to the main properties of the axion and its couplings.
We start by noticing the crucial role played by the coupling $\lambda_A$, in Eq. (\ref{V2}).
It follows from this term that the PQ charges of the scalar fields satisfy the relation 
\begin{eqnarray}\label{PQrel}
\frac{PQ_{\Phi_1}}{PQ_\sigma}+\frac{PQ_{\Phi_2}}{PQ_\sigma}+\frac{PQ_{\Phi_3}}{PQ_\sigma}= -1.
\end{eqnarray}
These charges can be written in terms of the scalar vevs, and when normalizing them by $PQ_\sigma$, we find 
\begin{eqnarray}\label{charges}
\frac{PQ_{\Phi_1}}{PQ_\sigma} = - \frac{v_2^2 w^2}{v_1^2v_2^2+v_{EW}^2w^2}, \quad\quad 
\frac{PQ_{\Phi_2}}{PQ_\sigma} = - \frac{v_1^2 w^2}{v_1^2v_2^2+v_{EW}^2w^2},\quad\quad
\frac{PQ_{\Phi_3}}{PQ_\sigma} = - \frac{v_1^2 v_2^2}{v_1^2v_2^2+v_{EW}^2w^2}.
\end{eqnarray}
Thus, we can use these charges to rewrite the axion profile in Eq. (\ref{axion}) in the usual form \cite{Srednicki:1985xd}
\begin{equation}\label{axion2}
 a = \frac{1}{f_{PQ}}\left[  v_1 PQ_{\Phi_1}\, a_1 +  v_2 PQ_{\Phi_2}\, a_2 +  w PQ_{\Phi_3}\, a_3 +  v_\sigma PQ_{\sigma}\,a_\sigma \right]\,,
\end{equation}
 where the dimensionful constant which normalizes the axion is defined as
\begin{eqnarray}\label{fPQ}
f_{PQ} &=& \sqrt{PQ_\sigma^2 v_\sigma^2+PQ_{\Phi_1}^2 v_1^2+PQ_{\Phi_2}^2 v_2^2+PQ_{\Phi_3}^2 w^2}\nonumber\\
&=& PQ_\sigma \sqrt{ v_\sigma^2 + \frac{v_1^2 v_2^2 w^2}{ v_1^2 v_2^2 +v_{EW}^2 w^2} }\,,
\end{eqnarray}  
where we have assumed that $ PQ_\sigma>0$. 

Notice that the axion decay constant, $f_a$, is in general defined as
\be \label{fa}
f_a \equiv \frac{f_{PQ}}{N_{DW}},
\ee
in terms of $f_{PQ}$, given in Eq. (\ref{fPQ}), and the domain wall number $N_{DW}$. In the present case, we have $N_{DW}=1$ so that the model is free from the domain wall problem.

Notice also that the PQ charges can be parametrized in terms of two angles:
\begin{eqnarray}
\frac{PQ_{\Phi_1}}{PQ_\sigma} =  - (\cos{\delta}\cos{\beta})^2,\quad\quad
\frac{PQ_{\Phi_2}}{PQ_\sigma} = - (\cos{\delta}\sin{\beta})^2,\quad\quad
\frac{PQ_{\Phi_3}}{PQ_\sigma} = - (\sin{\delta})^2,
\end{eqnarray}
where $\delta$ and $\beta$ are defined as
\begin{eqnarray}\label{angles}
\tan{\delta} = \frac{v_1v_2}{ w v_{EW}}\quad\quad\mbox{and} \quad\quad \tan{\beta} = \frac{v_1}{v_2}.
\end{eqnarray}
One sees that, as $\delta\to 0$ the axion has no $\Phi_3$ component and decouples from the exotic quarks.\\[-.2cm]  

Before turning to the discussion of axion couplings we mention the issue of the axion mass.
As usual, the axion field acquires a mass via nonpertubative QCD effects~\cite{Weinberg:1977ma,diCortona:2015ldu}
\begin{equation}
 m_a= \frac{\sqrt{m_u m_d}}{m_u + m_d}\frac{m_\pi f_\pi}{f_a} \simeq 5.7\left(\frac{10^{12}\,\mbox{GeV}}{f_a}\right) \mbox{$\mu$eV},
\end{equation}
with $m_u$, $m_d$, and $m_\pi$ the masses of the up quark, down quark and pion respectively, $f_\pi$ the pion decay constant, and $f_a$ is given in Eq. (\ref{fa}).
In the limit $v_\sigma\gg w \gg v_1,v_2$, it is easy to see that $a\simeq a_{\sigma}$, and the axion with $f_a\simeq v_\sigma$ is adequately ``invisibilized'' by the Dirac-neutrino seesaw scale.

It is well-known that the coherent oscillations of the axion field around its minimum may account for the cosmological cold dark matter~\cite{Abbott:1982af,Preskill:1982cy,Dine:1982ah}.
For $f_a\simeq v_\sigma$ in the range from $10^{9}-10^{12}\,\mbox{GeV}$, the axion can be relevant as cold dark matter in the usual manner, see~\cite{Sikivie:2020zpn,DiLuzio:2020wdo}.

\subsection{Standard model axion limits}
\label{sec:sm-axion-limits}

Let us now compare some of the main features of our construction with typical \sm invisible axion schemes. We recall that in the DFSZ models~\cite{Dine:1981rt,Zhitnitsky:1980tq} the standard fermions have tree-level coupling with the axion, since they carry $PQ$ charge. On the other hand, in the KSVZ models~\cite{Kim:1979if,Shifman:1979if} only the new fermions, with mass proportional to $v_\sigma$, carry $PQ$ charge, so that the axion there does not couple with \sm fermions. The crucial term in the comparison is
$$
\sigma\,\Phi_1\Phi_2\Phi_3
$$
which has no direct analogue within the standard model.
However, since the first two components of the triplets $\Phi_i$ form $\mathrm{SU(2)_L}$ doublets, $H_i$, and the third components, $\varphi_i$, are $\mathrm{SU(2)_L}$ singlets 
the above term would correspond in the \sm limit to 
$$\sigma \varphi_3 H_1 H_2.$$ 

The singlet $\sigma$ does not interact with fermions at tree-level and can be seen as the analogue of the Peccei-Quinn-charge-carrying singlet in the DFSZ models.
On the other hand, the third component of $\Phi_3$, $\varphi_3$, is similar to the scalar singlet in KSVZ models which, at tree-level, couples only to exotic fermions. 
Our model, therefore, can be understood as a hybrid DFSZ-KSVZ construction.
Taking our assumed vev hierarchy $v_\sigma\gg w\gg v_{EW}$, one sees from Eq. (\ref{charges}) that $PQ_{\Phi_3}/PQ_{\sigma}$, becomes suppressed.
This way we obtain the DFSZ-like limit.
It is also interesting to notice that our axion can only be ``invisibilized'' within this limit. \\[-.2cm]

In contrast to the original KSVZ model, in our proposal $\varphi_3$ not only plays a role in the breaking of the Peccei-Quinn symmetry, but is also responsible for the breaking of the extended electroweak gauge group characterizing the SVS theory.
Consequently, its CP-odd component, the field $a_3$, contributes mostly to the longitudinal Goldstone modes associated to $Z$ and $Z^\prime$ and
can not ``invisibilize'' the axion in a consistent manner.\\[-.2cm]

It follows that, at the standard \SM level, the viable invisible axion constructions would involve the $\sigma$ field, either through a quartic or a cubic term in the scalar potential,
$
\sigma^2\,H_1H_2~\mathrm{or}~\sigma\,H_1H_2.$ 

\subsection{Axion-to-photon coupling}
\label{sec:gauge-coupl-axion}

In order to ensure that the assumed $\mathrm{U(1)_{PQ}}$ symmetry of the model realizes the Peccei-Quinn mechanism for solving the strong $CP$ problem we must check that it produces an
  $[\mathrm{SU(3)_c}]^2 \times\mathrm{U(1)_{PQ}}$ anomaly. Indeed, the $\mathrm{U(1)_{PQ}}$ charges in Table~\ref{tab1} give the nonzero $[\mathrm{SU(3)_c}]^2 \mathrm{U(1)_{PQ}}$ anomaly coefficient
  $C_{ag}=PQ_\sigma$, as determined in Eq. (\ref{Cag}).  
As a result, one can turn the $\overline{\theta}$ parameter in the $CP$ violation term ${\cal L}\sim\overline{\theta}G\tilde{G}$ of the QCD Lagrangian into the dynamical axion field,
  which couples effectively to the gluon field strength, $G_{\mu\nu}^b$, according to 
\begin{eqnarray}
{\cal L}_{agg}=-\frac{\alpha_s}{8\pi}\frac{C_{ag}}{f_{PQ}}a\,G_{\mu\nu}^b\tilde{G}^{b,\mu\nu},
\end{eqnarray}
where $\tilde{G}^{b,\mu\nu}\equiv\epsilon^{\mu\nu\sigma\rho}G_{\sigma\rho}^b/2$ is the dual field strength, $\alpha_s=g_s/(4\pi)$ with $g_s$ the strong interaction coupling constant. \\[-.2cm]

Likewise the electromagnetic $[\mathrm{U(1)_Q}]^2\times \mathrm{U(1)_{PQ}}$ anomaly coefficient 
\begin{equation}\label{cagamma}
  C_{a\gamma} = 2 \sum_{i=charged}(PQ_{iL}-PQ_{iR})(Q^{i})^2
\end{equation}
is given by 
\begin{eqnarray}
  C_{a\gamma} &=&
  6 \left[ \left( 2 PQ_\sigma + 3 PQ_{\Phi_1} + 3 PQ_{\Phi_3} \right) \left( \frac{2}{3} \right)^2 - \left( PQ_\sigma + 3 PQ_{\Phi_1}+ 3 PQ_{\Phi_3} \right) \left( \frac{-1}{3} \right)^2  - \left( PQ_\sigma +  PQ_{\Phi_1} +  PQ_{\Phi_3} \right) (-1)^2 \right]\nonumber\\
  &=& -\frac{4}{3} PQ_\sigma\,.
\end{eqnarray}

The axion interaction with the electromagnetic field is dictated by the effective Lagrangian 
\be
{\cal L}_{a\gamma\gamma}=-\frac{g_{a\gamma}}{4}a\,F_{\mu\nu}\tilde{F}^{\mu\nu},
\label{laxga}
\ee
in which the axion-to-photon coupling is given as 
\be
g_{a\gamma}=\frac{\alpha}{2\pi f_a}\left(\frac{C_{a\gamma}}{C_{ag}}-\frac{2}{3}\frac{4+z}{1+z}\right)\approx\frac{\alpha}{2\pi f_a}\left(-\frac{4}{3}-1.95\right) \,,
\label{gag}
\ee
where $\alpha$ is the fine-structure constant and $z = m_u/m_d\approx 0.56$, the ratio between the up- and down-quark masses~\footnote{Further details on the derivation of the anomaly coefficients in the axion couplings with gluons, photons and fermions are given, for example, in~\cite{Dias:2014osa} and references therein. Note that the ratio of the anomaly coefficients $\frac{C_{a\gamma}}{C_{ag}}$ is commonly written as $\frac{E}{N}$ in the literature.}.
We stress that our prediction for the axion-to-photon coupling is a robust one, in the sense that it does not depend on the details of the Peccei-Quinn charge assignments made in Table~\ref{tab1}.
\\[-.2cm]  

  It is instructive to separate the coefficients in Eqs. (\ref{Cag}) and (\ref{cagamma}) into two contributions, one arising from the standard fermions ($st$),
  while the other comes exclusively from the exotic fermions ($ex$), as $C_{ag}=C_{ag}^{st}+C_{ag}^{ex}$ and $C_{a\gamma}=C_{a\gamma}^{st}+C_{a\gamma}^{ex}$, i.e.
\begin{eqnarray}\label{st+ex}
C_{ag}^{st} &=& PQ_\sigma + PQ_{\Phi_3}\quad \,\,\,\quad\quad \mbox{and} \quad C_{ag}^{ex} = - PQ_{\Phi_3},\\ 
C_{a\gamma}^{st} &=&-\frac{4}{3}\left( PQ_\sigma + PQ_{\Phi_3}\right)\quad \mbox{and}  \quad C_{a\gamma}^{ex} = \frac{4}{3} PQ_{\Phi_3}. \nonumber
\end{eqnarray}
In the limit where the exotic fermion contributions vanish, {\it i.e.} when $PQ_{\Phi_3}\to 0$, the predicted axion-photon coupling remains the same as in Eq. (\ref{gag}).
This could be understood as a ``DFSZ-like'' limit since only standard fermions contribute to the anomaly coefficients.
Nonetheless, instead of recovering the usual (flavor-universal) DFSZ constructions, we have a ``flavored'' axion as a result of the intrinsic flavor structure of 3-3-1 scenarios,
arising from the requirement of cancellation of the gauge anomalies~\cite{Singer:1980sw,Frampton:1992wt,Pisano:1991ee}.
This lies behind the different value we obtain for the $C_{a\gamma}/C_{ag}$ ratio when compared to the conventional standard-model-based flavor-universal axion schemes.\\[-.2cm]

Another interesting case, at least from the theory viewpoint, is the ``KSVZ-like'' limit, corresponding to $PQ_{\Phi_3}\to-PQ_{\sigma}$, achieved when $w\to 0$.
In this case only the exotic fermions contribute to the anomaly coefficients in Eq. (\ref{gag}).
This, however, would not be phenomenologically viable as the SVS new gauge bosons and exotic states would not acquire adequate masses.
This reinforces the discussion of Sec. \ref{sec:sm-axion-limits} where we found that the KSVZ-like limit of our model cannot be implemented in a consistent manner.

\subsection{Axion couplings to leptons}
\label{sec:charged-leptons}

The tree-level interactions between the axion and fermions can be obtained from the Yukawa sector. To find them, we use, in Eqs. (\ref{lagYq}) and (\ref{lagYl}), the profile of the axion, given in Eq. (\ref{axion2}), and rotate the fermions from the flavor to their mass bases, according to the adequate unitary transformations described in Secs. \ref{sec:quark-masses-mixing} and \ref{sec:lept-mass-mixings}, generically represented by
  $f_{L,R}\to U_{L,R}^{f}f^\prime_{L, R}$. Following this procedure for the charged leptons, we obtain
\begin{equation}\label{axioncl}
  -i g_{ae}\, a\, \overline{e^\prime} \gamma^5 e^\prime
\end{equation}
with
\begin{equation}
g_{ae} =\frac{\mbox{diag}(m_e,m_\mu,m_\tau)}{f_a} c_{ae}\quad\quad\mbox{and}\quad\quad  c_{ae}=\frac{C_{ae}}{C_{ag}}=\frac{PQ_{e_L}-PQ_{e_R}}{C_{ag}}=-\cos^2{\delta}\,\sin^2{\beta}, 
\end{equation}
where $e^\prime = e^\prime_L + e^\prime_R$. 

Notice that for values of the SVS scale consistent with 3-3-1 phenomenology, $w\gtrsim 10$ TeV, Eq. (\ref{angles}) implies that $\cos^2\delta\simeq 1$, so we have $c_{ae}\simeq -\sin^2 \beta$.
  This resembles the situation in the DFSZ model, except that our result is three times larger because of the domain-wall number.
  This similarity is expected since for $v_{EW}/w\ll 1$, our model leads to a low-energy two-Higgs-doublet effective axion model.\\[-.2cm]

  Turning to the axion couplings to the neutral leptons, in addition to diagonal contributions, we also expect non-diagonal terms.
This follows from the fact that the seesaw mechanism involves fields with different $\mathrm{U(1)_{PQ}}$ charges~\footnote{The situation is very much analogous
    to that characterizing the structure of the Majoron couplings within the conventional seesaw mechanism with spontaneous violation of lepton number~\cite{Schechter:1981cv}.}. 
The interaction terms can be written as, 
\begin{equation}\label{axionFCNCn}
     i a\, \overline{N^\prime_m}\left[(g_{aN}^V)_{mn}- (g_{aN}^A)_{mn}\gamma^5\right]N^\prime_n,
\end{equation}
with $m,n$ varying from $1$ to $6$, and $N^\prime$ representing the mass basis. As discussed above, this is related to the flavor basis via $N= U^N_{L,R}N^\prime$.
The vector and axial coefficients are given by
\begin{eqnarray}
(g_{aN}^V)_{mn} &=& \frac{m^N_m-m^N_n}{2 f_a}\left[(1+\cos^2\delta\cos^2\beta)\times X^{N_L}_{mn} - (1+ \sin^2\delta)\times X^{N_R}_{mn}\right],\\
(g_{aN}^A)_{mn} &=& \frac{m^N_m+m^N_n}{2 f_a}\left[(\cos^2\delta\sin^2\beta-2)\times \delta_{mn}+(1+\cos^2\delta\cos^2\beta)\times X^{N_L}_{mn} + (1+ \sin^2\delta)\times X^{N_R}_{mn}\right],\nonumber
\end{eqnarray}
with
\begin{equation}
 X^{N_{L,R}}_{mn} =\left[(U^N_{L,R})^\dagger\mbox{diag}\left(0_{3\times 3},I_{3\times 3}\right)U^N_{L,R}\right]_{mn}~.
\end{equation}
The $m_n^N$ are the eigenvalues of the neutral lepton mass matrix in Eq. (\ref{ssm}), {\it i.e.} the masses of both the active neutrinos $\nu^\prime$ and the heavy neutral mediator fermions $S^\prime$.

\subsection{Axion couplings to Quarks}
\label{sec:quarks}

When it comes to the axion couplings to standard quarks, the contributions are more involved as a result of the non-trivial embedding of quark families in different representations of
  $\mathrm{SU(3)_L}$, which leads to flavor changing neutral currents.
  This is a characteristic feature of 3-3-1 models and can be traced back to the cancellation of anomalies~\cite{Singer:1980sw,Frampton:1992wt,Pisano:1991ee}.
  The resulting axion-quark couplings are given as
\begin{equation}\label{axionFCNCq}
     i a\, \overline{q^\prime_i}\left[(g_{aq}^V)_{ij}- (g_{aq}^A)_{ij}\gamma^5\right]q^\prime_j,
\end{equation}
with $q^\prime = u^\prime, d^\prime$ and
\begin{eqnarray}
(g_{au}^V)_{ij} &=& \frac{m^u_i-m^u_j}{2 f_a}\cos^2\delta\times X^u_{ij},\quad\quad(g_{au}^A)_{ij} = \frac{m^u_i+m^u_j}{2 f_a}\cos^2\delta\left[\sin^2\beta \times \delta_{ij} + X^u_{ij}\right] ,\\
(g_{ad}^V)_{ij} &=& \frac{m^d_i-m^d_j}{2 f_a}\cos^2\delta\times X^d_{ij},\quad\quad(g_{ad}^A)_{ij} = \frac{m^d_i+m^d_j}{2 f_a}\cos^2\delta\left[\cos^2\beta \times \delta_{ij} + X^d_{ij}\right],\nonumber
\end{eqnarray}
with
\begin{equation}\label{eq:X}
 X^q_{ij} =\left[(U^q_L)^\dagger\mbox{diag}\left(0,0,-1\right)U^q_L\right]_{ij},  \quad\quad q = u,d,
\end{equation}
where $m^{u,d}_i$ are the masses of the up and down-type quarks, respectively.
An interesting and novel feature of our model is that, due to the non-standard embedding of quark families in $\mathrm{SU(3)_L}$ representations, one generically has flavour-changing axion couplings to quarks\footnote{ In contrast to the case of charged leptons, Eq. (\ref{axioncl}), the axion couples not only to axial but also vector quark currents.}.
These are encoded in the matrices $X_{ij}^{u,d}\not\propto \delta_{ij}$.
  Notice that flavour-changing axion couplings exist only for left-handed quarks, since right-handed ones all have the same PQ charge. This difference constitutes a structural feature of the theory.

For completeness we also give the axion couplings to the exotic quarks, $D_\alpha$ and $U_3$.
As in the case of the charged lepton in Eq. (\ref{axioncl}), the axion coupling to these fields are diagonal and can be described by the axial coefficients $g_{aD} =\mbox{diag}(m_{D_1},m_{D_2})\sin^2{\delta}/{f_a}$ and $g_{aU} =-m_{U}\sin^2{\delta}/{f_a}$. 

\subsection{Axion phenomenology}
\label{sec:axion-phenomenology}

We have now given the expressions for the axion couplings in our generalized axion scenario.
By embedding the axion in the extended \TrTrOO electroweak gauge symmetry we have encountered new features in the structure of these couplings. 
We now proceed to comment on their phenomenological implications, in particular on how they differ from existing axion models.

\begin{figure}[!h]
\begin{center}
\includegraphics[width=12cm,height=9cm]{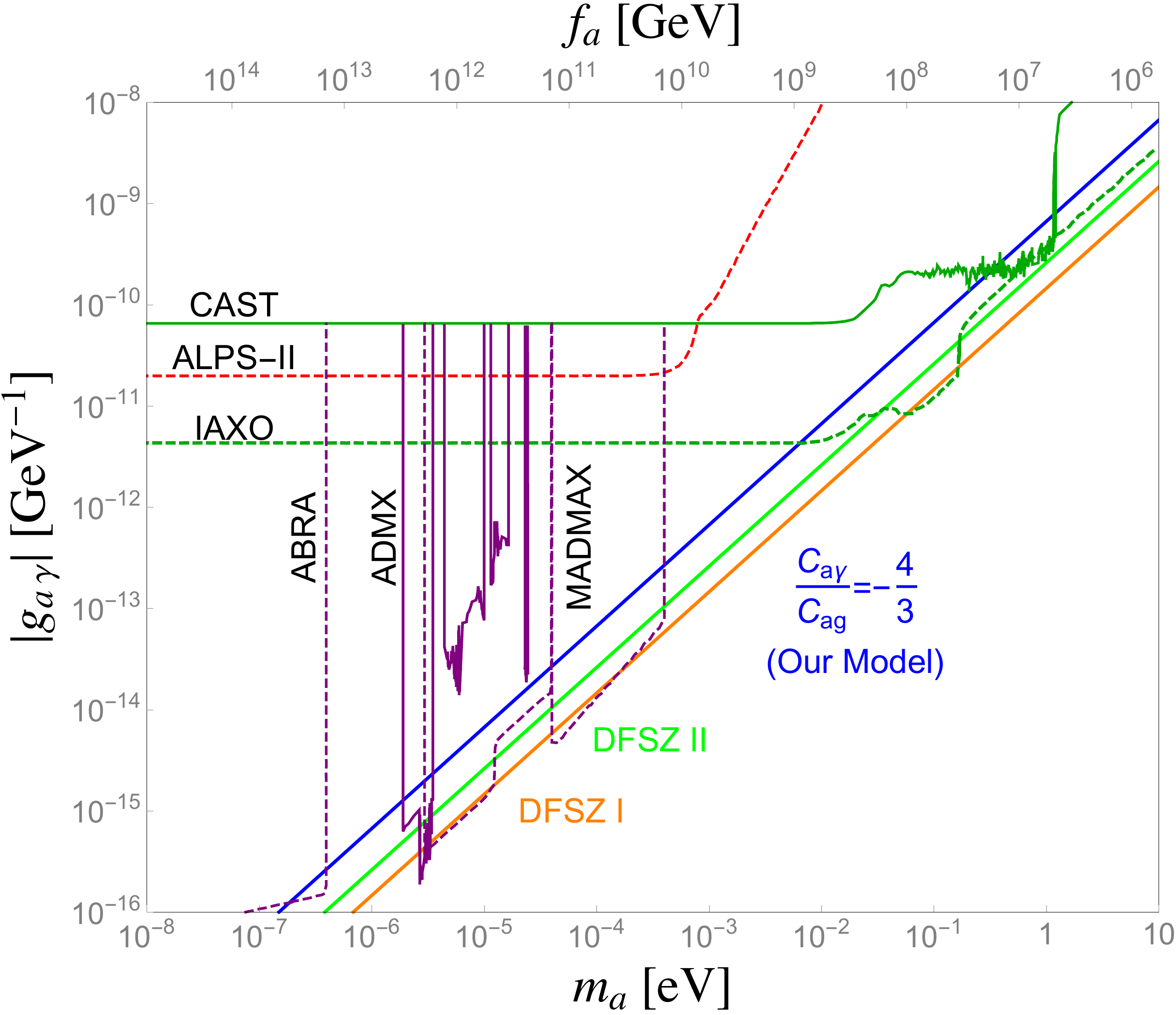} 
\caption{ Enhanced axion-to-photon coupling, compared to simplest DFSZ schemes. 
 The axion mass range from micro to mili-eV, relevant for dark matter, overlaps the sensitivities of the ADMX and MADMAX experiments.}
\label{plot:agg}
\end{center}
\end{figure}

There are several constraints on axion-photon coupling coming from laboratory searches, astrophysics and cosmology.
They have been recently compiled in Refs.~\cite{Sikivie:2020zpn,DiLuzio:2020wdo}. A summary is found in Fig.~\ref{plot:agg}.

We first discuss how models may be distinguished on the basis of their predicted value of $C_{a\gamma}$.  
We recall that there are two common versions of the DFSZ model, characterized by which one of the two Higgs doublets couples to the lepton fields~\cite{Cheng:1995fd}.
In the DFSZ-I model, such Higgs doublet is the same that couples to the down-quarks, so that the model has $\frac{C_{a\gamma}}{C_{ag}}=\frac{8}{3}$.
In the DFSZ-II model, the up-quarks and the leptons couple to the same Higgs doublet, so that in this model $\frac{C_{a\gamma}}{C_{ag}}=\frac{2}{3}$.
 
  In Fig.~\ref{plot:agg}, one sees that our proposed axion model predicts an enhanced value for $|g_{a\gamma}|$ when compared to the DFSZ-I and DFSZ-II models~\footnote{Note that axion predictions
    assume that the QCD anomaly is the only source of axion mass. In the presence of others, e.g. gravitational effects, the picture could change substantially. }.
This happens because our $\frac{C_{a\gamma}}{C_{ag}}$ has the same sign of the model independent part of the axion-photon coupling in Eq.~(\ref{gag}).
The larger predicted axion-to-photon coupling strength makes our axion lie within the expected sensitivities of the ABRACADABRA~\cite{Kahn:2016aff,Ouellet:2018beu}, ADMX~\cite{Duffy:2006aa,Asztalos:2009yp,Asztalos:2011bm,Stern:2016bbw,Braine:2019fqb}, MADMAX~\cite{TheMADMAXWorkingGroup:2016hpc} and IAXO~\cite{Vogel:2013bta,Armengaud:2019uso} experiments. 
Notice that this axion-to-photon coupling is a robust and uncontrived prediction of our model.
In principle, non-minimal \SM axion models containing extra Higgs doublets can also have the same $\frac{C_{a\gamma}}{C_{ag}}$ ratio as our model.
For example, the DFSZ-III axion model~\cite{Cheng:1995fd}, with three Higgs doublets, can have the same $\frac{C_{a\gamma}}{C_{ag}}$ ratio as our model.
Also models containing more than three Higgs doublets, such as the DFSZ-IV model, and/or quarks fields in exotic representations of the Standard Model gauge group
can produce values of $|g_{a\gamma}|$ larger than that of our model~\cite{DiLuzio:2017pfr}. 
Concerning models of the KSVZ-type, where the axion does not have tree-level couplings with the standard model fermions, it is interesting to notice that there are constructions with a larger
  value of $|g_{a\gamma}|$ compared to our model, but they feature domain-wall number $N_{DW}\geq 2$~\cite{DiLuzio:2016sbl,DiLuzio:2017pfr} (see also \cite{DiLuzio:2020wdo}).

\section{Conclusions and outlook}
\label{sec:Conclusions}

We have ``re-loaded'' the idea of the axion within an extension of the original SVS theory, using the electroweak \TrTrOO gauge symmetry.
This provides a comprehensive approach not only to the strong CP problem, but also to the existence of three families of fundamental fermions and the origin of neutrino masses.
Dark matter is axionic and directly related to the mechanism of neutrino mass generation.
Indeed, our proposed invisible axion theory leads to a type-I Dirac seesaw mechanism for neutrino masses, whose characteristic scale is set by the large Peccei-Quinn scale.
The observation of a positive neutrinoless double beta decay signal in this context would require some other physics,
for example, a short-range mechanism associated to additional scalar bosons beyond those in Table~\ref{tab1}~\cite{Schechter:1981bd}.

Let us stress that our construction differs from all previous implementations of the PQ mechanism within the 3-3-1 framework.
As an example we note that Ref.~\cite{Pal:1994ba} lacked a singlet PQ-carrying scalar, which plays a key role in making the axion invisible.
On the other hand, although in the 3-3-1 schemes of Refs.~\cite{Dias:2003zt,Dias:2003iq} neutrinos are Dirac particles, their small masses were not explained by a seesaw mechanism,
in contrast with our present model.
Finally, our proposal also differs from Ref.~\cite{Dias:2018ddy} where neutrinos get Majorana masses from a double seesaw mechanism.  
  
Our Dirac seesaw mechanism is suggestive of the existence of new physics, associated to the $\mathrm{SU(3)_L}$ gauge group, at a characteristic, relatively low scale, lying in between the weak scale and the PQ scale.
The model naturally leads to an enhanced axion coupling to photons, when compared with the simplest standard-model-based DFSZ-like models, see Fig. \ref{plot:agg}.
Moreover, in our scheme the couplings to fermions exhibit novel features, such as Eqs.~(\ref{axioncl}), (\ref{axionFCNCq}) and (\ref{eq:X}), which would deserve dedicated phenomenological study.

The phenomenological scope of our proposal is quite broad.
If the SVS scale is not too far above the electroweak scale, e.g. $w\gsim$ 10~TeV, as favored by the neutrino seesaw mechanism (see Eq.~(\ref{ssm2})), 
we expect di-lepton signatures from the production of the new $Z^\prime$ mediators through the Drell-Yan mechanism,
as well as flavor-changing effects in the decays of K, D and B mesons~\cite{Queiroz:2016gif}.
These would be a challenge both for high intensity as well as high energy experiments.
It is worthwhile to mention that besides dark matter and neutrino physics, the axion could be also connected with the cosmological inflation and 
baryogenesis~\cite{Ballesteros:2016euj,Ballesteros:2016xej}. 
Finally, we also comment that, in contrast to generic axion constructions, ours is free from the cosmological domain wall problem.

\black
\acknowledgements 
\noindent

Work supported by the Spanish grants FPA2017-85216-P (AEI/FEDER, UE), PROMETEO/2018/165 (Generalitat Valenciana) and the Spanish Red Consolider MultiDark FPA2017-90566-REDC. A. G. Dias thanks Conselho Nacional de Desenvolvimento Cient\'ifico e Tecnol\'ogico (CNPq) for its
financial support under the grant 305802/2019-4. J. L. acknowledges financial support under grants 2017/23027-2 and 2019/04195-7, S\~ao Paulo Research Foundation (FAPESP). 
CAV-A is supported by the Mexican C\'atedras CONACYT project 749 and SNI 58928.

\bibliographystyle{utphys}
\bibliography{bibliography}

\providecommand{\href}[2]{#2}\begingroup\raggedright\begin{thebibliography}{10}

\bibitem{Peccei:1977hh}
R.~Peccei and H.~R. Quinn, ``{CP Conservation in the Presence of Instantons},''
  \href{http://dx.doi.org/10.1103/PhysRevLett.38.1440}{{\em Phys.Rev.Lett.}
  {\bfseries 38} (1977) 1440--1443}.

\bibitem{Weinberg:1977ma}
S.~Weinberg, ``{A New Light Boson?},''
  \href{http://dx.doi.org/10.1103/PhysRevLett.40.223}{{\em Phys.Rev.Lett.}
  {\bfseries 40} (1978) 223--226}.

\bibitem{Wilczek:1977pj}
F.~Wilczek, ``{Problem of Strong $P$ and $T$ Invariance in the Presence of
  Instantons},'' \href{http://dx.doi.org/10.1103/PhysRevLett.40.279}{{\em
  Phys.Rev.Lett.} {\bfseries 40} (1978) 279--282}.

\bibitem{Dine:1981rt}
M.~Dine, W.~Fischler, and M.~Srednicki, ``{A Simple Solution to the Strong CP
  Problem with a Harmless Axion},''
  \href{http://dx.doi.org/10.1016/0370-2693(81)90590-6}{{\em Phys.Lett.}
  {\bfseries B104} (1981) 199--202}.

\bibitem{Zhitnitsky:1980tq}
A.~Zhitnitsky, ``{On Possible Suppression of the Axion Hadron Interactions. (In
  Russian)},'' {\em Sov.J.Nucl.Phys.} {\bfseries 31} (1980) 260.

\bibitem{Kim:1979if}
J.~E. Kim, ``{Weak Interaction Singlet and Strong CP Invariance},''
  \href{http://dx.doi.org/10.1103/PhysRevLett.43.103}{{\em Phys.Rev.Lett.}
  {\bfseries 43} (1979) 103}.

\bibitem{Shifman:1979if}
M.~A. Shifman, A.~Vainshtein, and V.~I. Zakharov, ``{Can Confinement Ensure
  Natural CP Invariance of Strong Interactions?},''
  \href{http://dx.doi.org/10.1016/0550-3213(80)90209-6}{{\em Nucl.Phys.}
  {\bfseries B166} (1980) 493--506}.

\bibitem{Sikivie:2020zpn}
P.~Sikivie, ``{Invisible Axion Search Methods},''
  \href{http://arxiv.org/abs/2003.02206}{{\ttfamily arXiv:2003.02206
  [hep-ph]}}.

\bibitem{DiLuzio:2020wdo}
L.~Di~Luzio, M.~Giannotti, E.~Nardi, and L.~Visinelli, ``{The landscape of QCD
  axion models},'' \href{http://arxiv.org/abs/2003.01100}{{\ttfamily
  arXiv:2003.01100 [hep-ph]}}.

\bibitem{Singer:1980sw}
M.~Singer, J.~W.~F. Valle, and J.~Schechter, ``{Canonical Neutral Current
  Predictions From the Weak Electromagnetic Gauge Group SU(3) X $u$(1)},''
\href{http://dx.doi.org/10.1103/PhysRevD.22.738}{{\em Phys. Rev.} {\bfseries
  D22} (1980) 738}.

\bibitem{Frampton:1992wt}
P.~Frampton, ``{Chiral dilepton model and the flavor question},''
  \href{http://dx.doi.org/10.1103/PhysRevLett.69.2889}{{\em Phys.Rev.Lett.}
  {\bfseries 69} (1992) 2889--2891}.

\bibitem{Pisano:1991ee}
F.~Pisano and V.~Pleitez, ``{An SU(3) x U(1) model for electroweak
  interactions},'' \href{http://dx.doi.org/10.1103/PhysRevD.46.410}{{\em
  Phys.Rev.} {\bfseries D46} (1992) 410--417}.

\bibitem{Byakti:2020ipa}
P.~Byakti and P.~B. Pal, ``{Generalized 331 models},''
  \href{http://arxiv.org/abs/2008.01266}{{\ttfamily arXiv:2008.01266
  [hep-ph]}}.

\bibitem{Boucenna:2015zwa}
S.~M. Boucenna, J.~W.~F. Valle, and A.~Vicente, ``{Predicting charged lepton
  flavor violation from 3-3-1 gauge symmetry},''
  \href{http://dx.doi.org/10.1103/PhysRevD.92.053001}{{\em Phys.Rev.}
  {\bfseries D92} (2015) 053001},
  \href{http://arxiv.org/abs/1502.07546}{{\ttfamily arXiv:1502.07546
  [hep-ph]}}.

\bibitem{deSalas:2020pgw}
P.~de~Salas {\em et~al.}, ``{2020 Global reassessment of the neutrino
  oscillation picture},'' \href{http://arxiv.org/abs/2006.11237}{{\ttfamily
  arXiv:2006.11237 [hep-ph]}}.

\bibitem{Bertone:2004pz}
G.~Bertone, D.~Hooper, and J.~Silk, ``{Particle dark matter: Evidence,
  candidates and constraints},''
  \href{http://dx.doi.org/10.1016/j.physrep.2004.08.031}{{\em Phys.Rept.}
  {\bfseries 405} (2005) 279--390}.

\bibitem{Abbott:1982af}
L.~Abbott and P.~Sikivie, ``{A Cosmological Bound on the Invisible Axion},''
  \href{http://dx.doi.org/10.1016/0370-2693(83)90638-X}{{\em Phys.Lett.}
  {\bfseries B120} (1983) 133--136}.

\bibitem{Preskill:1982cy}
J.~Preskill, M.~B. Wise, and F.~Wilczek, ``{Cosmology of the Invisible
  Axion},'' \href{http://dx.doi.org/10.1016/0370-2693(83)90637-8}{{\em
  Phys.Lett.} {\bfseries B120} (1983) 127--132}.

\bibitem{Dine:1982ah}
M.~Dine and W.~Fischler, ``{The Not So Harmless Axion},''
  \href{http://dx.doi.org/10.1016/0370-2693(83)90639-1}{{\em Phys.Lett.}
  {\bfseries B120} (1983) 137--141}.

\bibitem{deSalas:2017kay}
P.~de~Salas {\em et~al.}, ``{Status of neutrino oscillations 2018: 3$\sigma$
  hint for normal mass ordering and improved CP sensitivity},''
  \href{http://dx.doi.org/10.1016/j.physletb.2018.06.019}{{\em Phys.Lett.}
  {\bfseries B782} (2018) 633--640},
  \href{http://arxiv.org/abs/1708.01186}{{\ttfamily arXiv:1708.01186
  [hep-ph]}}.

\bibitem{Carena:2019nnd}
M.~Carena, D.~Liu, J.~Liu, N.~R. Shah, C.~E. Wagner, and X.-P. Wang, ``{$\nu$
  solution to the strong CP problem},''
  \href{http://dx.doi.org/10.1103/PhysRevD.100.094018}{{\em Phys.Rev.}
  {\bfseries D100} (2019) 094018},
  \href{http://arxiv.org/abs/1904.05360}{{\ttfamily arXiv:1904.05360
  [hep-ph]}}.

\bibitem{Dias:2005dn}
A.~G. Dias and V.~Pleitez, ``{The Invisible axion and neutrino masses},''
  \href{http://dx.doi.org/10.1103/PhysRevD.73.017701}{{\em Phys.Rev.}
  {\bfseries D73} (2006) 017701}.

\bibitem{Dasgupta:2013cwa}
B.~Dasgupta, E.~Ma, and K.~Tsumura, ``{Weakly interacting massive particle dark
  matter and radiative neutrino mass from Peccei-Quinn symmetry},''
  \href{http://dx.doi.org/10.1103/PhysRevD.89.041702}{{\em Phys.Rev.}
  {\bfseries D89} (2014) 041702},
  \href{http://arxiv.org/abs/1308.4138}{{\ttfamily arXiv:1308.4138 [hep-ph]}}.

\bibitem{Bertolini:2014aia}
S.~Bertolini, L.~Di~Luzio, H.~Kole{\v{s}}ov{\'a}, and M.~Malinsk{\'y},
  ``{Massive neutrinos and invisible axion minimally connected},''
  \href{http://dx.doi.org/10.1103/PhysRevD.91.055014}{{\em Phys.Rev.}
  {\bfseries D91} (2015) 055014},
  \href{http://arxiv.org/abs/1412.7105}{{\ttfamily arXiv:1412.7105 [hep-ph]}}.

\bibitem{Ahn:2015pia}
Y.~Ahn and E.~J. Chun, ``{Minimal Models for Axion and Neutrino},''
  \href{http://dx.doi.org/10.1016/j.physletb.2015.11.067}{{\em Phys.Lett.}
  {\bfseries B752} (2016) 333--337},
  \href{http://arxiv.org/abs/1510.01015}{{\ttfamily arXiv:1510.01015
  [hep-ph]}}.

\bibitem{Suematsu:2017kcu}
D.~Suematsu, ``{Dark matter stability and one-loop neutrino mass generation
  based on Peccei{\textendash}Quinn symmetry},''
  \href{http://dx.doi.org/10.1140/epjc/s10052-018-5519-4}{{\em Eur.Phys.J.}
  {\bfseries C78} (2018) 33}, \href{http://arxiv.org/abs/1709.02886}{{\ttfamily
  arXiv:1709.02886 [hep-ph]}}.

\bibitem{Ma:2017zyb}
E.~Ma, D.~Restrepo, and {\'O}.~Zapata, ``{Anomalous leptonic U(1) symmetry:
  Syndetic origin of the QCD axion, weak-scale dark matter, and radiative
  neutrino mass},'' \href{http://dx.doi.org/10.1142/S0217732318500244}{{\em
  Mod.Phys.Lett.} {\bfseries A33} (2018) 1850024},
  \href{http://arxiv.org/abs/1706.08240}{{\ttfamily arXiv:1706.08240
  [hep-ph]}}.

\bibitem{Reig:2018ocz}
M.~Reig, J.~W.~F. Valle, and F.~Wilczek, ``{SO(3) family symmetry and
  axions},'' \href{http://dx.doi.org/10.1103/PhysRevD.98.095008}{{\em
  Phys.Rev.} {\bfseries D98} (2018) 095008},
  \href{http://arxiv.org/abs/1805.08048}{{\ttfamily arXiv:1805.08048
  [hep-ph]}}.

\bibitem{Chen:2012baa}
C.-S. Chen and L.-H. Tsai, ``{Peccei-Quinn symmetry as the origin of Dirac
  Neutrino Masses},'' \href{http://dx.doi.org/10.1103/PhysRevD.88.055015}{{\em
  Phys.Rev.} {\bfseries D88} (2013) 055015},
  \href{http://arxiv.org/abs/1210.6264}{{\ttfamily arXiv:1210.6264 [hep-ph]}}.

\bibitem{Gu:2016hxh}
P.-H. Gu, ``{Peccei-Quinn symmetry for Dirac seesaw and leptogenesis},''
  \href{http://dx.doi.org/10.1088/1475-7516/2016/07/004}{{\em JCAP} {\bfseries
  1607} (2016) 004}, \href{http://arxiv.org/abs/1603.05070}{{\ttfamily
  arXiv:1603.05070 [hep-ph]}}.

\bibitem{Carvajal:2018ohk}
C.~D. Carvajal and {\'O}.~Zapata, ``{One-loop Dirac neutrino mass and mixed
  axion-WIMP dark matter},''
  \href{http://dx.doi.org/10.1103/PhysRevD.99.075009}{{\em Phys.Rev.}
  {\bfseries D99} (2019) 075009},
  \href{http://arxiv.org/abs/1812.06364}{{\ttfamily arXiv:1812.06364
  [hep-ph]}}.

\bibitem{Peinado:2019mrn}
E.~Peinado, M.~Reig, R.~Srivastava, and J.~W.~F. Valle, ``{Dirac neutrinos from
  Peccei-Quinn symmetry: a fresh look at the axion},''
  \href{http://arxiv.org/abs/1910.02961}{{\ttfamily arXiv:1910.02961
  [hep-ph]}}.

\bibitem{CentellesChulia:2018gwr}
S.~Centelles~Chuli{\'a}, R.~Srivastava, and J.~W.~F. Valle, ``{Seesaw roadmap
  to neutrino mass and dark matter},''
  \href{http://dx.doi.org/10.1016/j.physletb.2018.03.046}{{\em Phys. Lett.}
  {\bfseries B781} (2018) 122--128},
  \href{http://arxiv.org/abs/1802.05722}{{\ttfamily arXiv:1802.05722
  [hep-ph]}}.

\bibitem{CentellesChulia:2018bkz}
S.~Centelles~Chuli{\'a}, R.~Srivastava, and J.~W.~F. Valle, ``{Seesaw Dirac
  neutrino mass through dimension-six operators},''
  \href{http://dx.doi.org/10.1103/PhysRevD.98.035009}{{\em Phys.Rev.}
  {\bfseries D98} (2018) 035009},
  \href{http://arxiv.org/abs/1804.03181}{{\ttfamily arXiv:1804.03181
  [hep-ph]}}.

\bibitem{Chulia:2016giq}
S.~Centelles~Chuli{\'a}, R.~Srivastava, and J.~W.~F. Valle, ``{CP violation
  from flavor symmetry in a lepton quarticity dark matter model},''
  \href{http://dx.doi.org/10.1016/j.physletb.2016.08.028}{{\em Phys.Lett.}
  {\bfseries B761} (2016) 431--436},
  \href{http://arxiv.org/abs/1606.06904}{{\ttfamily arXiv:1606.06904
  [hep-ph]}}.

\bibitem{Bonilla:2016zef}
C.~Bonilla and J.~W.~F. Valle, ``{Naturally light neutrinos in $Diracon$
  model},'' \href{http://dx.doi.org/10.1016/j.physletb.2016.09.022}{{\em
  Phys.Lett.} {\bfseries B762} (2016) 162--165},
  \href{http://arxiv.org/abs/1605.08362}{{\ttfamily arXiv:1605.08362
  [hep-ph]}}.

\bibitem{Reig:2016ewy}
M.~Reig, J.~W.~F. Valle, and C.~Vaquera-Araujo, ``{Realistic $\mathrm{SU(3)_c
  \otimes SU(3)_L \otimes U(1)_X}$ model with a type II Dirac neutrino seesaw
  mechanism},'' \href{http://dx.doi.org/10.1103/PhysRevD.94.033012}{{\em
  Phys.Rev.} {\bfseries D94} (2016) 033012},
  \href{http://arxiv.org/abs/1606.08499}{{\ttfamily arXiv:1606.08499
  [hep-ph]}}.

\bibitem{Pal:1994ba}
P.~B. Pal, ``{The Strong CP question in SU(3)(C) x SU(3)(L) x U(1)(N)
  models},'' \href{http://dx.doi.org/10.1103/PhysRevD.52.1659}{{\em Phys.Rev.}
  {\bfseries D52} (1995) 1659--1662}.

\bibitem{Dias:2003zt}
A.~G. Dias and V.~Pleitez, ``{Stabilizing the invisible axion in 3-3-1
  models},'' \href{http://dx.doi.org/10.1103/PhysRevD.69.077702}{{\em
  Phys.Rev.} {\bfseries D69} (2004) 077702}.

\bibitem{Dias:2003iq}
A.~G. Dias, C.~A. de~S.~Pires, and P.~S. Rodrigues~da Silva, ``{Discrete
  symmetries, invisible axion and lepton number symmetry in an economic 3 3 1
  model},'' \href{http://dx.doi.org/10.1103/PhysRevD.68.115009}{{\em Phys.Rev.}
  {\bfseries D68} (2003) 115009}.

\bibitem{Dias:2018ddy}
A.~Dias, J.~Leite, D.~Lopes, and C.~Nishi, ``{Fermion Mass Hierarchy and Double
  Seesaw Mechanism in a 3-3-1 Model with an Axion},''
  \href{http://dx.doi.org/10.1103/PhysRevD.98.115017}{{\em Phys.Rev.}
  {\bfseries D98} (2018) 115017},
  \href{http://arxiv.org/abs/1810.01893}{{\ttfamily arXiv:1810.01893
  [hep-ph]}}.

\bibitem{Dong:2013wca}
P.~Dong, H.~Hung, and T.~Tham, ``{3-3-1-1 model for dark matter},''
  \href{http://dx.doi.org/10.1103/PhysRevD.87.115003}{{\em Phys.Rev.}
  {\bfseries D87} (2013) 115003},
  \href{http://arxiv.org/abs/1305.0369}{{\ttfamily arXiv:1305.0369 [hep-ph]}}.

\bibitem{Leite:2020bnb}
J.~Leite, A.~Morales, J.~W.~F. Valle, and C.~A. Vaquera-Araujo, ``{Dark matter
  stability from Dirac neutrinos in scotogenic 3-3-1-1 theory},''
  \href{http://dx.doi.org/10.1103/PhysRevD.102.015022}{{\em Phys. Rev. D}
  {\bfseries 102} no.~1, (2020) 015022},
  \href{http://arxiv.org/abs/2005.03600}{{\ttfamily arXiv:2005.03600
  [hep-ph]}}.

\bibitem{Valle:1983dk}
J.~W.~F. Valle and M.~Singer, ``{Lepton Number Violation With Quasi Dirac
  Neutrinos},'' \href{http://dx.doi.org/10.1103/PhysRevD.28.540}{{\em Phys.
  Rev. D} {\bfseries 28} (1983) 540}.

\bibitem{Schechter:1981cv}
J.~Schechter and J.~W.~F. Valle, ``Neutrino decay and spontaneous violation of
  lepton number,'' \href{http://dx.doi.org/10.1103/PhysRevD.25.774}{{\em Phys.
  Rev. D} {\bfseries 25} (Feb, 1982) 774--783}.
  \url{https://link.aps.org/doi/10.1103/PhysRevD.25.774}.

\bibitem{Schechter:1980gr}
J.~Schechter and J.~W.~F. Valle, ``{Neutrino Masses in SU(2) x U(1)
  Theories},'' \href{http://dx.doi.org/10.1103/PhysRevD.22.2227}{{\em
  Phys.Rev.} {\bfseries D22} (1980) 2227}.

\bibitem{Rodejohann:2011vc}
W.~Rodejohann and J.~W.~F. Valle, ``{Symmetrical Parametrizations of the Lepton
  Mixing Matrix},'' \href{http://dx.doi.org/10.1103/PhysRevD.84.073011}{{\em
  Phys.Rev.} {\bfseries D84} (2011) 073011},
  \href{http://arxiv.org/abs/1108.3484}{{\ttfamily arXiv:1108.3484 [hep-ph]}}.

\bibitem{Srednicki:1985xd}
M.~Srednicki, ``{Axion Couplings to Matter. 1. CP Conserving Parts},''
  \href{http://dx.doi.org/10.1016/0550-3213(85)90054-9}{{\em Nucl.Phys.}
  {\bfseries B260} (1985) 689--700}.

\bibitem{diCortona:2015ldu}
G.~Grilli~di Cortona, E.~Hardy, J.~Pardo~Vega, and G.~Villadoro, ``{The QCD
  axion, precisely},'' \href{http://dx.doi.org/10.1007/JHEP01(2016)034}{{\em
  JHEP} {\bfseries 1601} (2016) 034},
  \href{http://arxiv.org/abs/1511.02867}{{\ttfamily arXiv:1511.02867
  [hep-ph]}}.

\bibitem{Dias:2014osa}
A.~Dias, A.~Machado, C.~Nishi, A.~Ringwald, and P.~Vaudrevange, ``{The Quest
  for an Intermediate-Scale Accidental Axion and Further ALPs},''
  \href{http://dx.doi.org/10.1007/JHEP06(2014)037}{{\em JHEP} {\bfseries 06}
  (2014) 037}, \href{http://arxiv.org/abs/1403.5760}{{\ttfamily arXiv:1403.5760
  [hep-ph]}}.

\bibitem{Cheng:1995fd}
S.~Cheng, C.~Geng, and W.~Ni, ``{Axion - photon couplings in invisible axion
  models},'' \href{http://dx.doi.org/10.1103/PhysRevD.52.3132}{{\em Phys. Rev.
  D} {\bfseries 52} (1995) 3132--3135},
  \href{http://arxiv.org/abs/hep-ph/9506295}{{\ttfamily arXiv:hep-ph/9506295}}.

\bibitem{Kahn:2016aff}
Y.~Kahn, B.~R. Safdi, and J.~Thaler, ``{Broadband and Resonant Approaches to
  Axion Dark Matter Detection},''
  \href{http://dx.doi.org/10.1103/PhysRevLett.117.141801}{{\em Phys.Rev.Lett.}
  {\bfseries 117} (2016) 141801},
  \href{http://arxiv.org/abs/1602.01086}{{\ttfamily arXiv:1602.01086
  [hep-ph]}}.

\bibitem{Ouellet:2018beu}
J.~L. Ouellet {\em et~al.}, ``{First Results from ABRACADABRA-10 cm: A Search
  for Sub-$\mu$eV Axion Dark Matter},''
  \href{http://dx.doi.org/10.1103/PhysRevLett.122.121802}{{\em Phys.Rev.Lett.}
  {\bfseries 122} (2019) 121802},
  \href{http://arxiv.org/abs/1810.12257}{{\ttfamily arXiv:1810.12257
  [hep-ex]}}.

\bibitem{Duffy:2006aa}
{\bfseries ADMX} Collaboration, L.~D. Duffy {\em et~al.}, ``{A high resolution
  search for dark-matter axions},''
  \href{http://dx.doi.org/10.1103/PhysRevD.74.012006}{{\em Phys.Rev.}
  {\bfseries D74} (2006) 012006}.

\bibitem{Asztalos:2009yp}
{\bfseries ADMX} Collaboration, S.~Asztalos {\em et~al.}, ``{A SQUID-based
  microwave cavity search for dark-matter axions},''
  \href{http://dx.doi.org/10.1103/PhysRevLett.104.041301}{{\em Phys.Rev.Lett.}
  {\bfseries 104} (2010) 041301},
  \href{http://arxiv.org/abs/0910.5914}{{\ttfamily arXiv:0910.5914
  [astro-ph.CO]}}.

\bibitem{Asztalos:2011bm}
{\bfseries ADMX} Collaboration, S.~Asztalos {\em et~al.}, ``{Design and
  performance of the ADMX SQUID-based microwave receiver},''
  \href{http://dx.doi.org/10.1016/j.nima.2011.07.019}{{\em Nucl.Instrum.Meth.}
  {\bfseries A656} (2011) 39--44},
  \href{http://arxiv.org/abs/1105.4203}{{\ttfamily arXiv:1105.4203
  [physics.ins-det]}}.

\bibitem{Stern:2016bbw}
I.~Stern, \href{http://dx.doi.org/10.22323/1.282.0198}{``{ADMX Status},''}
  vol.~ICHEP2016, p.~198.
\newblock 2016.
\newblock \href{http://arxiv.org/abs/1612.08296}{{\ttfamily arXiv:1612.08296
  [physics.ins-det]}}.

\bibitem{Braine:2019fqb}
{\bfseries ADMX} Collaboration, T.~Braine {\em et~al.}, ``{Extended Search for
  the Invisible Axion with the Axion Dark Matter Experiment},''
  \href{http://dx.doi.org/10.1103/PhysRevLett.124.101303}{{\em Phys.Rev.Lett.}
  {\bfseries 124} (2020) 101303},
  \href{http://arxiv.org/abs/1910.08638}{{\ttfamily arXiv:1910.08638
  [hep-ex]}}.

\bibitem{TheMADMAXWorkingGroup:2016hpc}
{\bfseries MADMAX Working Group} Collaboration, A.~Caldwell {\em et~al.},
  ``{Dielectric Haloscopes: A New Way to Detect Axion Dark Matter},''
  \href{http://dx.doi.org/10.1103/PhysRevLett.118.091801}{{\em Phys.Rev.Lett.}
  {\bfseries 118} (2017) 091801},
  \href{http://arxiv.org/abs/1611.05865}{{\ttfamily arXiv:1611.05865
  [physics.ins-det]}}.

\bibitem{Vogel:2013bta}
J.~Vogel {\em et~al.}, ``{IAXO - The International Axion Observatory},''
\newblock 2013.
\newblock \href{http://arxiv.org/abs/1302.3273}{{\ttfamily arXiv:1302.3273
  [physics.ins-det]}}.

\bibitem{Armengaud:2019uso}
{\bfseries IAXO} Collaboration, E.~Armengaud {\em et~al.}, ``{Physics potential
  of the International Axion Observatory (IAXO)},''
  \href{http://dx.doi.org/10.1088/1475-7516/2019/06/047}{{\em JCAP} {\bfseries
  1906} (2019) 047}, \href{http://arxiv.org/abs/1904.09155}{{\ttfamily
  arXiv:1904.09155 [hep-ph]}}.

\bibitem{DiLuzio:2017pfr}
L.~Di~Luzio, F.~Mescia, and E.~Nardi, ``{Window for preferred axion models},''
  \href{http://dx.doi.org/10.1103/PhysRevD.96.075003}{{\em Phys. Rev. D}
  {\bfseries 96} no.~7, (2017) 075003},
  \href{http://arxiv.org/abs/1705.05370}{{\ttfamily arXiv:1705.05370
  [hep-ph]}}.

\bibitem{DiLuzio:2016sbl}
L.~Di~Luzio, F.~Mescia, and E.~Nardi, ``{Redefining the Axion Window},''
  \href{http://dx.doi.org/10.1103/PhysRevLett.118.031801}{{\em Phys.Rev.Lett.}
  {\bfseries 118} (2017) 031801},
  \href{http://arxiv.org/abs/1610.07593}{{\ttfamily arXiv:1610.07593
  [hep-ph]}}.

\bibitem{Schechter:1981bd}
J.~Schechter and J.~W.~F. Valle, ``{Neutrinoless Double beta Decay in SU(2) x
  U(1) Theories},'' \href{http://dx.doi.org/10.1103/PhysRevD.25.2951}{{\em
  Phys.Rev.} {\bfseries D25} (1982) 2951}.

\bibitem{Queiroz:2016gif}
F.~S. Queiroz, C.~Siqueira, and J.~W.~F. Valle, ``{Constraining Flavor Changing
  Interactions from LHC Run-2 Dilepton Bounds with Vector Mediators},''
  \href{http://dx.doi.org/10.1016/j.physletb.2016.10.057}{{\em Phys.Lett.}
  {\bfseries B763} (2016) 269--274},
  \href{http://arxiv.org/abs/1608.07295}{{\ttfamily arXiv:1608.07295
  [hep-ph]}}.

\bibitem{Ballesteros:2016euj}
G.~Ballesteros, J.~Redondo, A.~Ringwald, and C.~Tamarit, ``{Unifying inflation
  with the axion, dark matter, baryogenesis and the seesaw mechanism},''
  \href{http://dx.doi.org/10.1103/PhysRevLett.118.071802}{{\em Phys.Rev.Lett.}
  {\bfseries 118} (2017) 071802},
  \href{http://arxiv.org/abs/1608.05414}{{\ttfamily arXiv:1608.05414
  [hep-ph]}}.

\bibitem{Ballesteros:2016xej}
G.~Ballesteros, J.~Redondo, A.~Ringwald, and C.~Tamarit, ``{Standard
  Model{\textemdash}axion{\textemdash}seesaw{\textemdash}Higgs portal
  inflation. Five problems of particle physics and cosmology solved in one
  stroke},'' \href{http://dx.doi.org/10.1088/1475-7516/2017/08/001}{{\em JCAP}
  {\bfseries 1708} (2017) 001},
  \href{http://arxiv.org/abs/1610.01639}{{\ttfamily arXiv:1610.01639
  [hep-ph]}}.

\end{thebibliography}\endgroup
\end{document}